\documentclass[aps,superscriptaddress,nofootinbib,floatfix,notitlepage,twocolumn]{revtex4-2}
\usepackage[utf8x]{inputenc}
\pdfoutput=1
\usepackage{graphicx}
\usepackage{hyperref}
\usepackage{bm}
\usepackage{multirow}
\usepackage{array,booktabs,colortbl,multirow}
\usepackage{colortbl,xcolor,colordvi,color}
\usepackage{rotating}
\usepackage{placeins}
\usepackage{amsmath}

\newcolumntype{C}[1]{>{\centering\let\newline\\\arraybackslash\hspace{0pt}}m{#1}}

\usepackage[sort&compress]{natbib}

\begin{document}

\title{Impact of the recent measurements of the top-quark and \texorpdfstring{$\mathbf{W}$-boson}{W boson} masses on electroweak precision fits}

\author{J.~de Blas}
\affiliation{CAFPE and Departamento de F\'isica Te\'orica y del Cosmos, Universidad de Granada, Campus de Fuentenueva, E–18071 Granada, Spain}

\author{M.~Pierini}
\affiliation{CERN, 1211 Geneva 23, Switzerland}

\author{L.~Reina}
\affiliation{Physics Department, Florida State University,\\ Tallahassee, FL 32306-4350, USA}

\author{L.~Silvestrini}
\affiliation{INFN, Sezione di Roma, Piazzale A. Moro 2, I-00185 Roma, Italy}


\begin{abstract}
  We assess the impact of the very recent measurement of the top-quark
  mass by the CMS Collaboration \cite{CMS-PAS-TOP-20-008} on the fit of electroweak
  data in the Standard Model and beyond, with particular emphasis on
  the prediction for the mass of the $W$ boson. We then compare this
  prediction with the average of the corresponding experimental measurements
  including the
  new measurement by the CDF Collaboration
  \cite{CDF:2022hxs}, and discuss its compatibility in the Standard Model, in new
  physics models with oblique corrections, and in the dimension-six 
  Standard Model Effective Field Theory. Finally, we present the updated global fit to electroweak precision data in these models.
\end{abstract}
\maketitle

\begin{figure*}[htb]
  \centering
  \includegraphics[width=.425\textwidth]{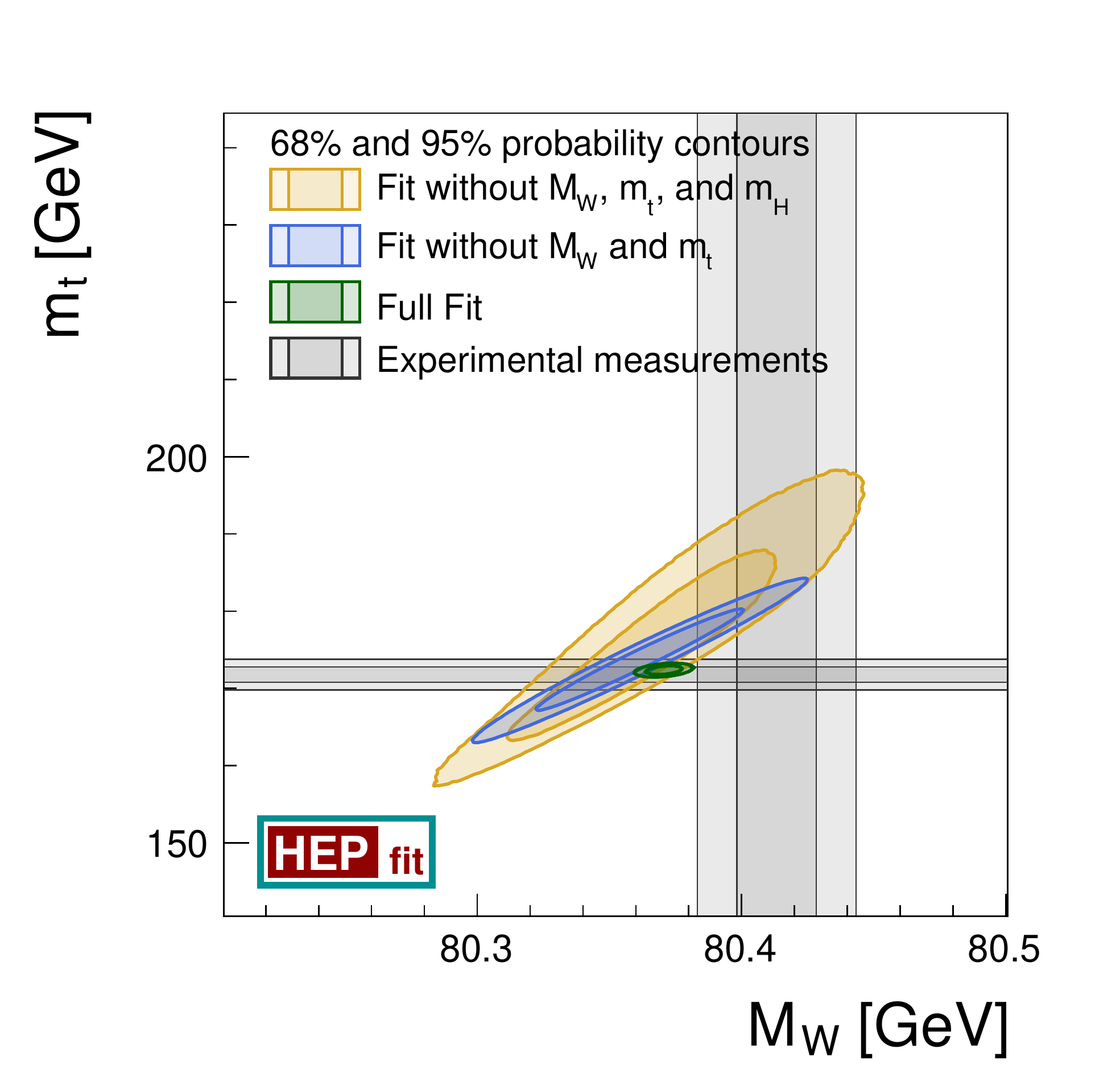}
  \includegraphics[width=.425\textwidth]{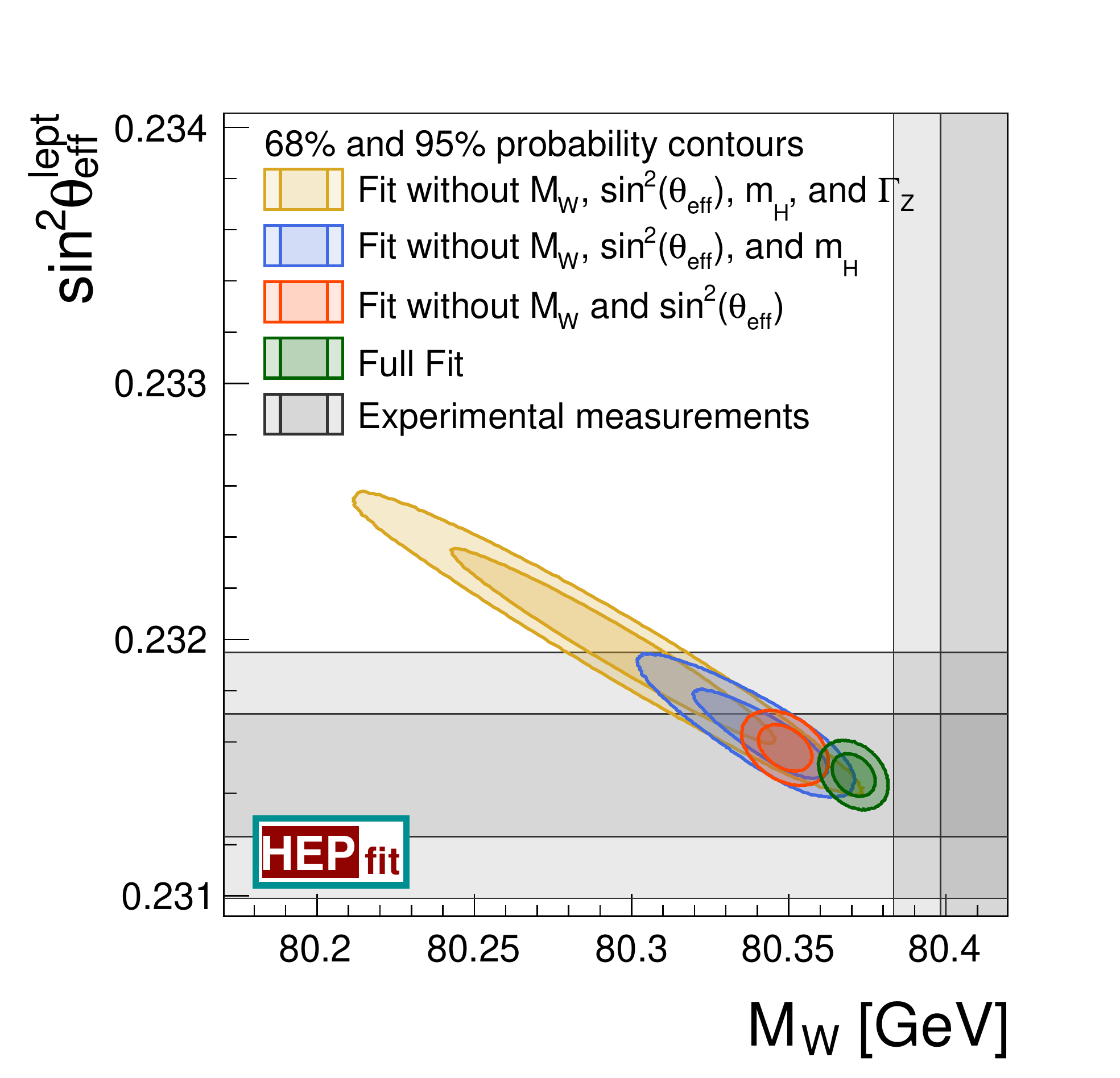}
  \caption{Posterior from a global fit of all EWPO in the SM in the $m_t$ vs. $M_W$ (top) and $\sin^2\theta_{\rm eff}^{\rm lept}$ vs. $M_W$ (bottom) planes, superimposed to the posteriors obtained omitting different observables from the fit in the \emph{standard average} scenario. Dark (light) regions
    correspond to $68\%$ ($95\%$) probability ranges. Direct measurements are shown in grey. The corresponding results in the \emph{conservative average} scenario are presented in Figure \protect\ref{fig:mtmwsm_C}.}
  \label{fig:mtmwsm}
\end{figure*}

The mass of the top quark ($m_t$) plays a crucial role in the study of
Standard Model (SM) predictions for precision observables in the
ElectroWeak (EW) and flavour sectors, since several amplitudes are
quadratically sensitive to $m_t$. Indeed, indirect bounds on the
top-quark mass were obtained using EW and flavour observables well before
its direct measurement \cite{Ellis:1977uk,Buras:1992uf}. Nowadays, $m_t$ gives the
dominant parametric uncertainty on several EW Precision Observables
(EWPO) \cite{deBlas:2021wap}, among which is the $W$-boson mass
($M_W$). The posterior from a global fit omitting or including the experimental
information on $m_t$ and $M_W$ is reported in
Figure~\ref{fig:mtmwsm}. (We also show in the
same figure analogous information in the 
$\sin^2{\theta_{\rm eff}^{\rm lept}}$ vs. $M_W$ plane.)
All posteriors reported in this paper are
obtained from a Bayesian analysis performed with the {\tt HEPfit} code
\cite{deBlas:2019okz}, using state-of-the-art calculations for all
EWPO\footnote{A thorough description of all elements entering the EWPO global fit used in this Letter is given in Ref.~\cite{deBlas:2021wap}, to which we refer the reader interested in such details.} \cite{Sirlin:1980nh,Marciano:1980pb,Djouadi:1987gn,Djouadi:1987di,Kniehl:1989yc,Halzen:1990je,Kniehl:1991gu,Kniehl:1992dx,Barbieri:1992nz,Barbieri:1992dq,Djouadi:1993ss,Fleischer:1993ub,Fleischer:1994cb,Avdeev:1994db,Chetyrkin:1995ix,Chetyrkin:1995js,Degrassi:1996mg,Degrassi:1996ps,Degrassi:1999jd,Freitas:2000gg,vanderBij:2000cg,Freitas:2002ja,Awramik:2002wn,Onishchenko:2002ve,Awramik:2002vu,Awramik:2002wv,Awramik:2003ee,Awramik:2003rn,Faisst:2003px,Dubovyk:2016aqv,Dubovyk:2018rlg}. All inputs used are reported in Table \ref{tab:SM_std}, while the theory uncertainties we use are:
\begin{eqnarray}
    \label{eq:therr}
    &&\delta_\mathrm{th} M_W = 4\, \mathrm{MeV}\,,\quad \delta_\mathrm{th} \sin^2{\theta_W} = 5\cdot 10^{-5}\,,\\
    &&\delta_\mathrm{th} \Gamma_{Z} = 0.4\, \mathrm{MeV} \,,\quad
    \delta_\mathrm{th} \sigma^0_\mathrm{had} = 6\, \mathrm{pb} \,,\quad \nonumber \\
    &&\delta_\mathrm{th} R^0_\ell = 0.006 \,,\quad \delta_\mathrm{th} R^0_c = 0.00005 \,,\quad
    \delta_\mathrm{th} R^0_b = 0.0001 \,.\quad \nonumber
\end{eqnarray}

From Figure~\ref{fig:mtmwsm} it is evident that $m_t$ and $M_W$ are
tightly correlated in the SM, so that experimental improvements in either one might
challenge the validity of the SM and provide us with precious hints on
what kind of New Physics (NP) might be present at yet unprobed energy
scales. Indeed, this is precisely the situation once the very recent
measurement of $m_t$ from the CMS Collaboration
\cite{CMS-PAS-TOP-20-008},
\begin{equation}
  \label{eq:CMSmt}
 m_t = 171.77 \pm 0.38~\mathrm{GeV}\,, 
\end{equation}
and of $M_W$
from the CDF Collaboration \cite{CDF:2022hxs},
\begin{equation}
  \label{eq:CDFMW}
  M_W = 80.4335 \pm 0.0094~\mathrm{GeV}\,,
\end{equation}
are
included in the analysis. This Letter is dedicated to assessing the
impact of these measurements in the SM and in several parametrizations
of physics beyond the SM.

Let us first consider the impact of the new measurement of $m_t$ in Eq.~(\ref{eq:CMSmt}). Following Ref.~\cite{deBlas:2021wap}, we combine the 2016
Tevatron combination~\cite{TevatronElectroweakWorkingGroup:2016lid}, the 2015
CMS Run 1 combination~\cite{Khachatryan:2015hba}, the combination of ATLAS Run 1 results in Ref.~\cite{Aaboud:2018zbu},
the CMS Run 2 measurements in the dilepton, lepton+jets, all-jet and single-top channels \cite{CMS-PAS-TOP-20-008,Sirunyan:2018goh,Sirunyan:2018mlv,CMS:2021jnp}
and the ATLAS Run 2 result from the lepton+jet channel~\cite{ATLAS:2019ezb},  assuming the linear correlation coefficient between two systematic uncertainties to be written as $\rho_{ij}^\mathrm{sys}=\mathrm{min}\left\{\sigma_i^\mathrm{sys}, \sigma_j^\mathrm{sys}\right\}/\mathrm{max}\left\{\sigma_i^\mathrm{sys}, \sigma_j^\mathrm{sys}\right\}$. In this way we obtain a new average (compared to Ref.~\cite{deBlas:2021wap}) given by:
\begin{equation}
  \label{eq:mtave_std}
  m_t = 171.79 \pm 0.38~\mathrm{GeV}\,,
\end{equation}
where the uncertainty is dominated, as expected, by the very recent CMS measurement. However, since this average does not take into account the tensions between individual measurements, we also consider a \emph{conservative average} in which the error is inflated to $1$ GeV. 

For the $W$-boson mass, we compute the average of all the existing measurements from LEP 2, the Tevatron, and the LHC. The new measurement from CDF gives, when combined with the D0 one, a Tevatron combination of ($80.427\pm 0.0089$)~GeV~\cite{CDF:2022hxs}. This was combined with the LHC ATLAS~\cite{ATLAS:2017rzl} and LHCb \cite{LHCb:2021bjt} measurements assuming a common systematic uncertainty of 4.7~MeV, corresponding to the CDF uncertainty from PDF and QED radiation. The resulting number is combined in an uncorrelated manner with the LEP2 determination, obtaining as new average:\footnote{We observe that the result of the combination does not depend strongly on the value of the common uncertainty between 0 and 6.9~MeV, the total CDF systematic uncertainty~\cite{CDF:2022hxs}. In particular, the combined uncertainty ranges between 7.7 and 8.4~MeV, whereas the central values can change by slightly less than 1~\!$\sigma$. Thus, waiting for an official combination of LHC and TeVatron results, we take the result in Eq.~\protect\ref{eq:MWave_std} as our best estimate of $M_W$.}
\begin{equation}
    M_W=80.4133\pm 0.0080~\mathrm{GeV}.
    \label{eq:MWave_std}
\end{equation}
As in the top-quark mass case, there is however a significant tension between the new CDF measurement and the other measurements that enter in the calculation of Eq.~(\ref{eq:MWave_std}), with $\chi^2/n_{\mathrm{dof}}=3.59$. Therefore, in a \textit{conservative average}, we inflate the error on $M_W$ to 0.015~GeV.

We then perform a series of fits to the different EWPO using both the {\it standard} (see Eqs.~\ref{eq:mtave_std} and \ref{eq:MWave_std}) and {\it conservative} assumptions for the uncertainties of the top-quark and $W$-boson masses.~\footnote{Unlike in Ref.~\cite{deBlas:2021wap}, we do not consider an inflated uncertainty for the Higgs-boson mass in the  {\it conservative} scenario since, as noted in that reference, this has little impact on the output of the EW fit. We thus use $m_H=(125.21\pm 0.12$)~GeV in all the fits presented here.}
(Although we will discuss both scenarios throughout the text, in most of the tables and figures in the main text we will report the results pertaining to the {\it standard average}. The results for the {\it conservative average} scenario are shown in the appendix.)
In particular, we are interested in comparing the new averages with the corresponding predictions obtained in the SM. For that purpose we first perform a pure SM fit of all EWPO, excluding the experimental input for $M_W$ and, from the posterior of such fit, we compute the SM prediction for $M_W$. The results are shown in Table~\ref{tab:MW_fits}, where we also compare with the combined $M_W$ values in each scenario via the 1D pull, computed as explained in Ref.~\cite{deBlas:2021wap}. As it is apparent, there exists a significant 6.5~\!$\sigma$  discrepancy with the SM in the \emph{standard average}, which persists at the level of 3.7~\!$\sigma$ even in the \emph{conservative} scenario, due to the large difference between the new CDF measurement and the SM prediction. 
\begin{table}[htb]
    \centering
    \begin{tabular}{c|cc|cc}
    \hline
        Model & Pred. $M_W$ [GeV] & Pull & Pred. $M_W$ [GeV] & Pull  \\
         & \multicolumn{2}{c|}{\emph{standard average}} &  \multicolumn{2}{c}{\emph{conservative average}} \\
         \hline
         SM &  $80.3499 \pm 0.0056$ & $\phantom{+}6.5\,\sigma$ & $80.3505 \pm 0.0077$ & $\phantom{+}3.7\,\sigma$ \\
         ST &  $80.366 \pm 0.029$ & $\phantom{+}1.6\,\sigma$ &  $80.367 \pm 0.029$ & $\phantom{+}1.4\,\sigma$  \\
         STU &  $80.32 \pm 0.54 $ & $\phantom{+}0.2\,\sigma$ &  $80.32 \pm 0.54  $ & $\phantom{+}0.2\,\sigma$  \\         
         SMEFT & $80.66 \pm 1.68$ & $-0.1\,\sigma$ &  $80.66 \pm 1.68$ & $-0.1\,\sigma$  \\
         \hline
    \end{tabular}
    \caption{Predictions and pulls for $M_W$ in the SM, in the \textit{oblique} NP models and in the SMEFT, using the \emph{standard} and \emph{conservative} averaging scenarios. The predictions are obtained without using the experimental information on $M_W$. See text for more details.}
    \label{tab:MW_fits}
\end{table}

\begin{table*}[htbp]
    \centering
    \resizebox{\textwidth}{!}{
    \begin{tabular}{l|c|c|cr|cr|cr}
\toprule
& Measurement & Posterior & Indirect/Prediction & Pull & Full Indirect & Pull & Full Prediction & Pull \\
\hline 
$\alpha_{s}(M_{Z})$ & $ 0.1177 \pm 0.0010 $ & $ 0.11762 \pm 0.00095 $ & $ 0.11685 \pm 0.00278 $ & $ 0.3 $ & $ 0.12181 \pm 0.00470 $ & $ -0.8 $ & $ 0.1177 \pm 0.0010 $ &$ - $ \\ 
& & $[ 0.11576 , 0.11946 ]$ & $[ 0.11145 , 0.12233 ]$ & & $[ 0.1126 , 0.1310 ]$ & & $[ 0.1157 , 0.1197 ]$  \\ 
$\Delta\alpha^{(5)}_{\mathrm{had}}(M_Z)$ & $ 0.02766 \pm 0.00010 $ & $ 0.027535 \pm 0.000096 $ & $ 0.026174 \pm 0.000334 $ & $ 4.3 $ & $ 0.028005 \pm 0.000675 $ & $ -0.5 $ & $ 0.02766 \pm 0.00010 $ &$ - $ \\ 
& & $[ 0.027349 , 0.027726 ]$ & $[ 0.025522 , 0.026826 ]$ & & $[ 0.02667 , 0.02932 ]$ & & $[ 0.02746 , 0.02786 ]$  \\ 
$M_Z$ [GeV] & $ 91.1875 \pm 0.0021 $ & $ 91.1911 \pm 0.0020 $ & $ 91.2314 \pm 0.0069 $ & $ -6.1 $ & $ 91.2108 \pm 0.0390 $ & $ -0.6 $ & $ 91.1875 \pm 0.0021 $ &$ - $ \\ 
& & $[ 91.1872 , 91.1950 ]$ & $[ 91.2178 , 91.2447 ]$ & & $[ 91.136 , 91.288 ]$ & & $[ 91.1834 , 91.1916 ]$  \\ 
$m_t$ [GeV] & $ 171.79 \pm 0.38 $ & $ 172.36 \pm 0.37 $ & $ 181.45 \pm 1.49 $ & $ -6.3 $ & $ 187.58 \pm 9.52 $ & $ -1.7 $ & $ 171.80 \pm 0.38 $ &$ - $ \\ 
& & $[ 171.64 , 173.09 ]$ & $[ 178.53 , 184.42 ]$ & & $[ 169.1 , 206.1 ]$ & & $[ 171.05 , 172.54 ]$  \\ 
$m_H$ [GeV] & $ 125.21 \pm 0.12 $ & $ 125.20 \pm 0.12 $ & $ 93.36 \pm 4.99 $ & $ 4.3 $ & $ 247.98 \pm 125.35 $ & $ -0.9 $ & $ 125.21 \pm 0.12 $ &$ - $ \\ 
& & $[ 124.97 , 125.44 ]$ & $[ 82.92 , 102.89 ]$ & & $[ 100.8 , 640.4 ]$ & & $[ 124.97 , 125.45 ]$  \\ 
\hline 
$M_W$ [GeV] & $ 80.4133 \pm 0.0080 $ & $ 80.3706 \pm 0.0045 $ & $ 80.3499 \pm 0.0056 $ & $ 6.5 $ & $ 80.4129 \pm 0.0080 $ & $ 0.1 $ & $ 80.3496 \pm 0.0057 $ &$ 6.5 $ \\ 
& & $[ 80.3617 , 80.3794 ]$ & $[ 80.3391 , 80.3610 ]$ & & $[ 80.3973 , 80.4284 ]$ & & $[ 80.3386 , 80.3608 ]$  \\ 
\hline 
$\Gamma_{W}$ [GeV] & $ 2.085 \pm 0.042 $ & $ 2.08903 \pm 0.00053 $ & $ 2.08902 \pm 0.00052 $ & $ -0.1 $ & $ 2.09430 \pm 0.00224 $ & $ -0.2 $ & $ 2.08744 \pm 0.00059 $ &$ 0.0 $ \\ 
& & $[ 2.08800 , 2.09006 ]$ & $[ 2.08799 , 2.09005 ]$ & & $[ 2.0900 , 2.0988 ]$ & & $[ 2.08627 , 2.08859 ]$  \\ 
\hline 
$\sin^2\theta_{\rm eff}^{\rm lept}(Q_{\rm FB}^{\rm had})$ & $ 0.2324 \pm 0.0012 $ & $ 0.231471 \pm 0.000055 $ & $ 0.231469 \pm 0.000056 $ & $ 0.8 $ & $ 0.231460 \pm 0.000138 $ & $ 0.8 $ & $ 0.231558 \pm 0.000062 $ &$ 0.7 $ \\ 
& & $[ 0.231362 , 0.231580 ]$ & $[ 0.231361 , 0.231578 ]$ & & $[ 0.23119 , 0.23173 ]$ & & $[ 0.231436 , 0.231679 ]$  \\ 
\hline 
$P_{\tau}^{\rm pol}=\mathcal{A}_\ell$ & $ 0.1465 \pm 0.0033 $ & $ 0.14742 \pm 0.00044 $ & $ 0.14744 \pm 0.00044 $ & $ -0.3 $ & $ 0.14750 \pm 0.00108 $ & $ -0.3 $ & $ 0.14675 \pm 0.00049 $ &$ -0.1 $ \\ 
& & $[ 0.14656 , 0.14827 ]$ & $[ 0.14657 , 0.14830 ]$ & & $[ 0.1454 , 0.1496 ]$ & & $[ 0.14580 , 0.14770 ]$  \\ 
\hline 
$\Gamma_{Z}$ [GeV] & $ 2.4955 \pm 0.0023 $ & $ 2.49455 \pm 0.00065 $ & $ 2.49437 \pm 0.00068 $ & $ 0.5 $ & $ 2.49530 \pm 0.00204 $ & $ 0.0 $ & $ 2.49397 \pm 0.00068 $ &$ 0.6 $ \\ 
& & $[ 2.49329 , 2.49581 ]$ & $[ 2.49301 , 2.49569 ]$ & & $[ 2.4912 , 2.4993 ]$ & & $[ 2.49262 , 2.49531 ]$  \\ 
$\sigma_{h}^{0}$ [nb] & $ 41.480 \pm 0.033 $ & $ 41.4892 \pm 0.0077 $ & $ 41.4914 \pm 0.0080 $ & $ -0.3 $ & $ 41.4613 \pm 0.0303 $ & $ 0.4 $ & $ 41.4923 \pm 0.0080 $ &$ -0.4 $ \\ 
& & $[ 41.4741 , 41.5041 ]$ & $[ 41.4757 , 41.5070 ]$ & & $[ 41.402 , 41.521 ]$ & & $[ 41.4766 , 41.5081 ]$  \\ 
$R^{0}_{\ell}$ & $ 20.767 \pm 0.025 $ & $ 20.7487 \pm 0.0080 $ & $ 20.7451 \pm 0.0087 $ & $ 0.8 $ & $ 20.7587 \pm 0.0217 $ & $ 0.2 $ & $ 20.7468 \pm 0.0087 $ &$ 0.7 $ \\ 
& & $[ 20.7329 , 20.7645 ]$ & $[ 20.7281 , 20.7621 ]$ & & $[ 20.716 , 20.801 ]$ & & $[ 20.7298 , 20.7637 ]$  \\ 
$A_{\rm FB}^{0, \ell}$ & $ 0.0171 \pm 0.0010 $ & $ 0.016300 \pm 0.000095 $ & $ 0.016291 \pm 0.000096 $ & $ 0.8 $ & $ 0.016316 \pm 0.000240 $ & $ 0.8 $ & $ 0.01615 \pm 0.00011 $ &$ 1.0 $ \\ 
& & $[ 0.016111 , 0.016487 ]$ & $[ 0.016102 , 0.016480 ]$ & & $[ 0.01585 , 0.01679 ]$ & & $[ 0.01594 , 0.01636 ]$  \\ 
\hline 
$\mathcal{A}_{\ell}$ (SLD) & $ 0.1513 \pm 0.0021 $ & $ 0.14742 \pm 0.00044 $ & $ 0.14745 \pm 0.00045 $ & $ 1.8 $ & $ 0.14750 \pm 0.00108 $ & $ 1.6 $ & $ 0.14675 \pm 0.00049 $ &$ 2.1 $ \\ 
& & $[ 0.14656 , 0.14827 ]$ & $[ 0.14656 , 0.14834 ]$ & & $[ 0.1454 , 0.1496 ]$ & & $[ 0.14580 , 0.14770 ]$  \\ 
$R^{0}_{b}$ & $ 0.21629 \pm 0.00066 $ & $ 0.215892 \pm 0.000100 $ & $ 0.215886 \pm 0.000102 $ & $ 0.6 $ & $ 0.215413 \pm 0.000364 $ & $ 1.2 $ & $ 0.21591 \pm 0.00010 $ &$ 0.6 $ \\ 
& & $[ 0.215696 , 0.216089 ]$ & $[ 0.215688 , 0.216086 ]$ & & $[ 0.21469 , 0.21611 ]$ & & $[ 0.21571 , 0.21611 ]$  \\ 
$R^{0}_{c}$ & $ 0.1721 \pm 0.0030 $ & $ 0.172198 \pm 0.000054 $ & $ 0.172197 \pm 0.000054 $ & $ -0.1 $ & $ 0.172404 \pm 0.000183 $ & $ -0.1 $ & $ 0.172189 \pm 0.000054 $ &$ -0.1 $ \\ 
& & $[ 0.172093 , 0.172302 ]$ & $[ 0.172094 , 0.172303 ]$ & & $[ 0.17206 , 0.17278 ]$ & & $[ 0.172084 , 0.172295 ]$  \\ 
$A_{\rm FB}^{0, b}$ & $ 0.0996 \pm 0.0016 $ & $ 0.10335 \pm 0.00030 $ & $ 0.10337 \pm 0.00032 $ & $ -2.3 $ & $ 0.10338 \pm 0.00077 $ & $ -2.1 $ & $ 0.10288 \pm 0.00034 $ &$ -2.0 $ \\ 
& & $[ 0.10276 , 0.10396 ]$ & $[ 0.10275 , 0.10400 ]$ & & $[ 0.10189 , 0.10490 ]$ & & $[ 0.10220 , 0.10354 ]$  \\ 
$A_{\rm FB}^{0, c}$ & $ 0.0707 \pm 0.0035 $ & $ 0.07385 \pm 0.00023 $ & $ 0.07387 \pm 0.00023 $ & $ -0.9 $ & $ 0.07392 \pm 0.00059 $ & $ -0.9 $ & $ 0.07348 \pm 0.00025 $ &$ -0.8 $ \\ 
& & $[ 0.07341 , 0.07430 ]$ & $[ 0.07341 , 0.07434 ]$ & & $[ 0.07275 , 0.07507 ]$ & & $[ 0.07298 , 0.07398 ]$  \\ 
$\mathcal{A}_b$ & $ 0.923 \pm 0.020 $ & $ 0.934770 \pm 0.000039 $ & $ 0.934772 \pm 0.000040 $ & $ -0.6 $ & $ 0.934593 \pm 0.000166 $ & $ -0.6 $ & $ 0.934721 \pm 0.000041 $ &$ -0.6 $ \\ 
& & $[ 0.934693 , 0.934847 ]$ & $[ 0.934693 , 0.934849 ]$ & & $[ 0.93426 , 0.93491 ]$ & & $[ 0.934642 , 0.934801 ]$  \\ 
$\mathcal{A}_c$ & $ 0.670 \pm 0.027 $ & $ 0.66796 \pm 0.00021 $ & $ 0.66797 \pm 0.00021 $ & $ 0.1 $ & $ 0.66817 \pm 0.00054 $ & $ 0.1 $ & $ 0.66766 \pm 0.00022 $ &$ 0.1 $ \\ 
& & $[ 0.66754 , 0.66838 ]$ & $[ 0.66755 , 0.66839 ]$ & & $[ 0.66712 , 0.66922 ]$ & & $[ 0.66722 , 0.66810 ]$  \\ 
\hline 
$\mathcal{A}_s$ & $ 0.895 \pm 0.091 $ & $ 0.935678 \pm 0.000039 $ & $ 0.935677 \pm 0.000040 $ & $ -0.4 $ & $ 0.935716 \pm 0.000098 $ & $ -0.5 $ & $ 0.935621 \pm 0.000041 $ &$ -0.5 $ \\ 
& & $[ 0.935600 , 0.935755 ]$ & $[ 0.935599 , 0.935754 ]$ & & $[ 0.935523 , 0.935909 ]$ & & $[ 0.935541 , 0.935702 ]$  \\ 
BR$_{W\to\ell\bar\nu_\ell}$ & $ 0.10860 \pm 0.00090 $ & $ 0.108388 \pm 0.000022 $ & $ 0.108388 \pm 0.000022 $ & $ 0.2 $ & $ 0.108291 \pm 0.000109 $ & $ 0.3 $ & $ 0.108386 \pm 0.000023 $ &$ 0.2 $ \\ 
& & $[ 0.108345 , 0.108431 ]$ & $[ 0.108345 , 0.108431 ]$ & & $[ 0.10808 , 0.10851 ]$ & & $[ 0.108340 , 0.108432 ]$  \\ 
$\sin^2\theta_{\rm eff}^{\rm lept}$ (HC) & $ 0.23143 \pm 0.00025 $ & $ 0.231471 \pm 0.000055 $ & $ 0.231474 \pm 0.000056 $ & $ -0.2 $ & $ 0.231460 \pm 0.000138 $ & $ -0.1 $ & $ 0.231558 \pm 0.000062 $ &$ -0.5 $ \\ 
& & $[ 0.231362 , 0.231580 ]$ & $[ 0.231363 , 0.231584 ]$ & & $[ 0.23119 , 0.23173 ]$ & & $[ 0.231436 , 0.231679 ]$  \\ 
\hline 
$R_{uc}$ & $ 0.1660 \pm 0.0090 $ & $ 0.172220 \pm 0.000031 $ & $ 0.172220 \pm 0.000032 $ & $ -0.7 $ & $ 0.172424 \pm 0.000180 $ & $ -0.7 $ & $ 0.172212 \pm 0.000032 $ &$ -0.7 $ \\ 
& & $[ 0.172159 , 0.172282 ]$ & $[ 0.172159 , 0.172282 ]$ & & $[ 0.17209 , 0.17279 ]$ & & $[ 0.172149 , 0.172275 ]$  \\ 
 \bottomrule
    \end{tabular}}
    \caption{Experimental data, Posterior from the full fit, Indirect determination of individual SM paramers/Prediction of individual EWPO, Full Indirect determination of all SM parameters simultaneously, and Full Prediction of all EWPO simultaneously in the \emph{standard average} scenario. The (Full) Indirect determination/(Full) Prediction is obtained omitting the experimental information on individual (all) SM parameters/individual (all) EWPO.}
    \label{tab:SM_std}
\end{table*}

In Tables \ref{tab:SM_std} and \ref{tab:SM_c} we present, in addition to the experimental values for all EWPO used, the posterior from the global fit, the prediction of individual parameters/observables obtained omitting the corresponding experimental information, the indirect determination of SM parameters obtained solely from EWPO, and the full prediction obtained using only the experimental information on SM parameters. For the individual prediction, indirect determination and for the full prediction we also report the pull for each experimental result. 
In this regard, from the individual indirect determination of the SM parameters in Table \ref{tab:SM_std}, one can observe how the tensions introduced by the new measurements in the SM fit result into sizable pulls for the different SM inputs, at the level of 4~\!$\sigma$ (6~\!$\sigma$) for $\Delta\alpha^{(5)}_{\mathrm{had}}(M_Z)$ and $m_H$ ($M_Z$ and $m_t$). 
Each pull can be converted in a \textit{p-value}, and the global consistency of the SM in the EWPO domain can be tested by looking at the distribution of \textit{p-values}. From Table \ref{tab:SM_std}, in the indirect determination case, we find an average \textit{p-value} of $0.43$ with a $0.36$ standard deviation, while for the full prediction we obtain an average \textit{p-value} of $0.56$ with a $0.30$ standard deviation. Both values are compatible with the expectation of a flatly distributed \textit{p-value} between zero and one. Furthermore, we evaluate the global \textit{p-value} from the full prediction, taking into account all theoretical and experimental correlations. We obtain $p=2.45\cdot 10^{-5}$, corresponding to a global pull of $4.2\,\sigma$, in the \emph{standard} averaging scenario, and $p=0.10$, corresponding to a global pull of $1.6\,\sigma$, in the \emph{conservative} averaging scenario.
 
\begin{figure*}[htb]
    \centering
    \includegraphics[width=0.245\textwidth]{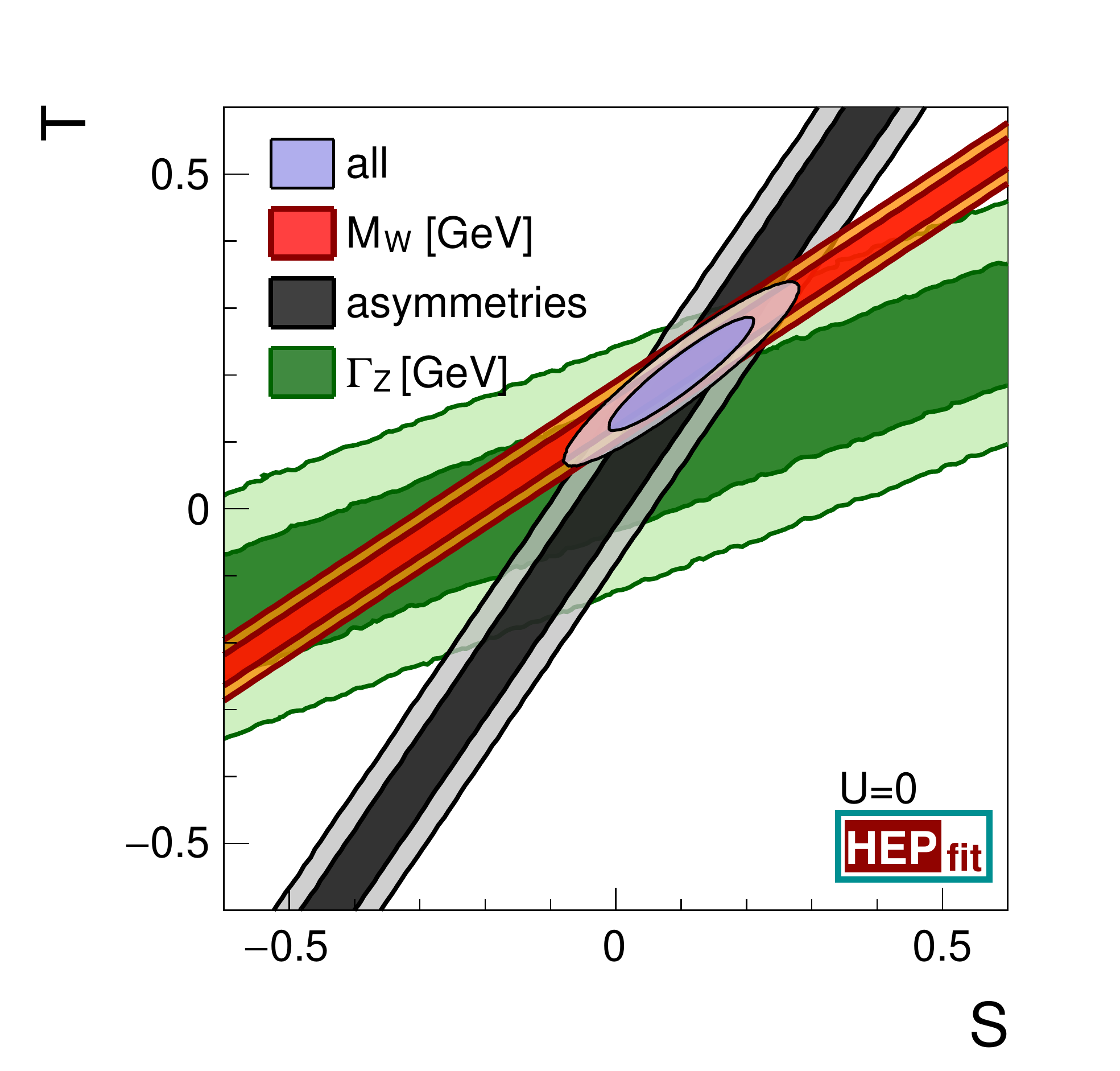}
    \includegraphics[width=0.245\textwidth]{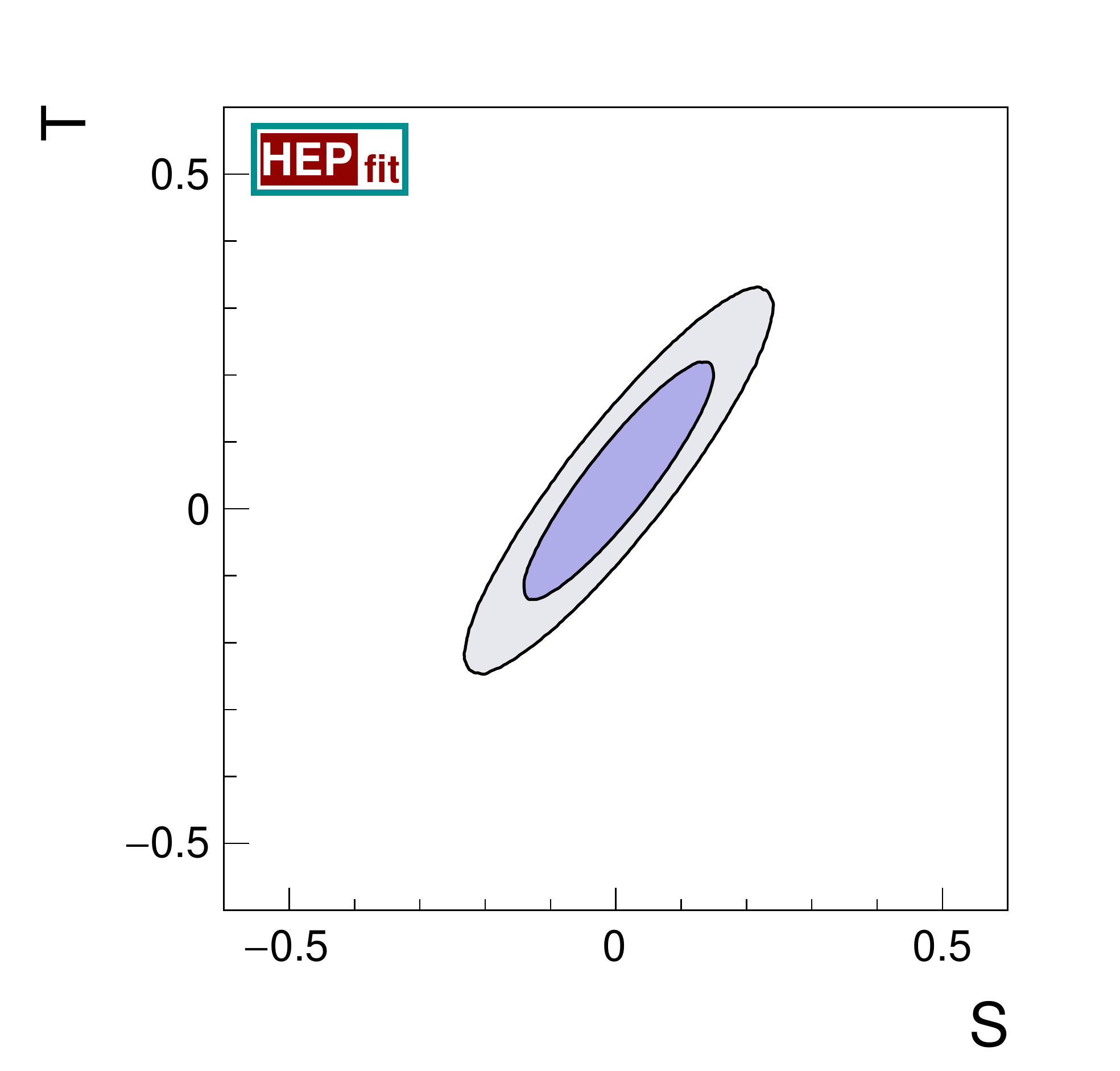}
    \includegraphics[width=0.245\textwidth]{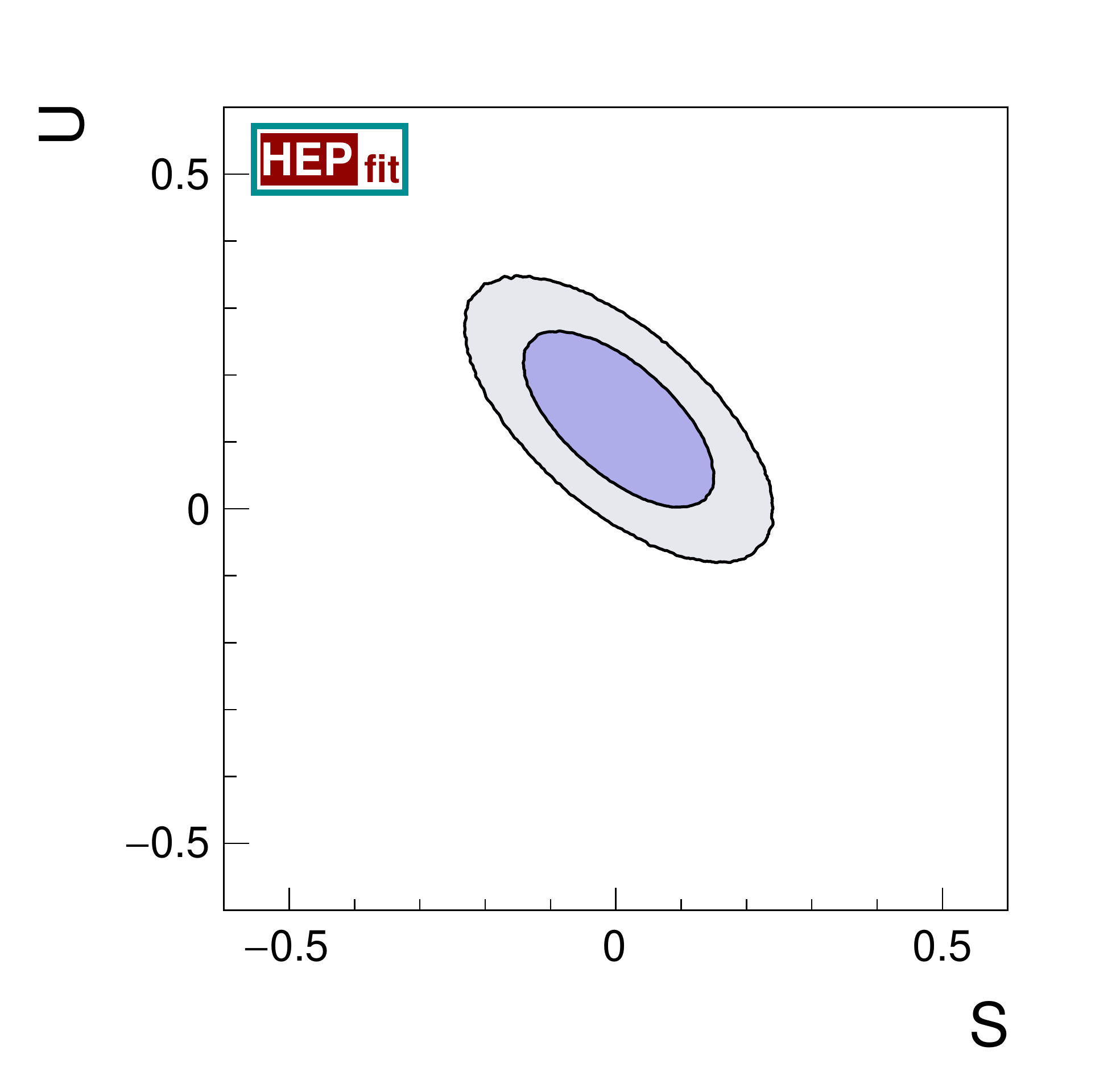}
    \includegraphics[width=0.245\textwidth]{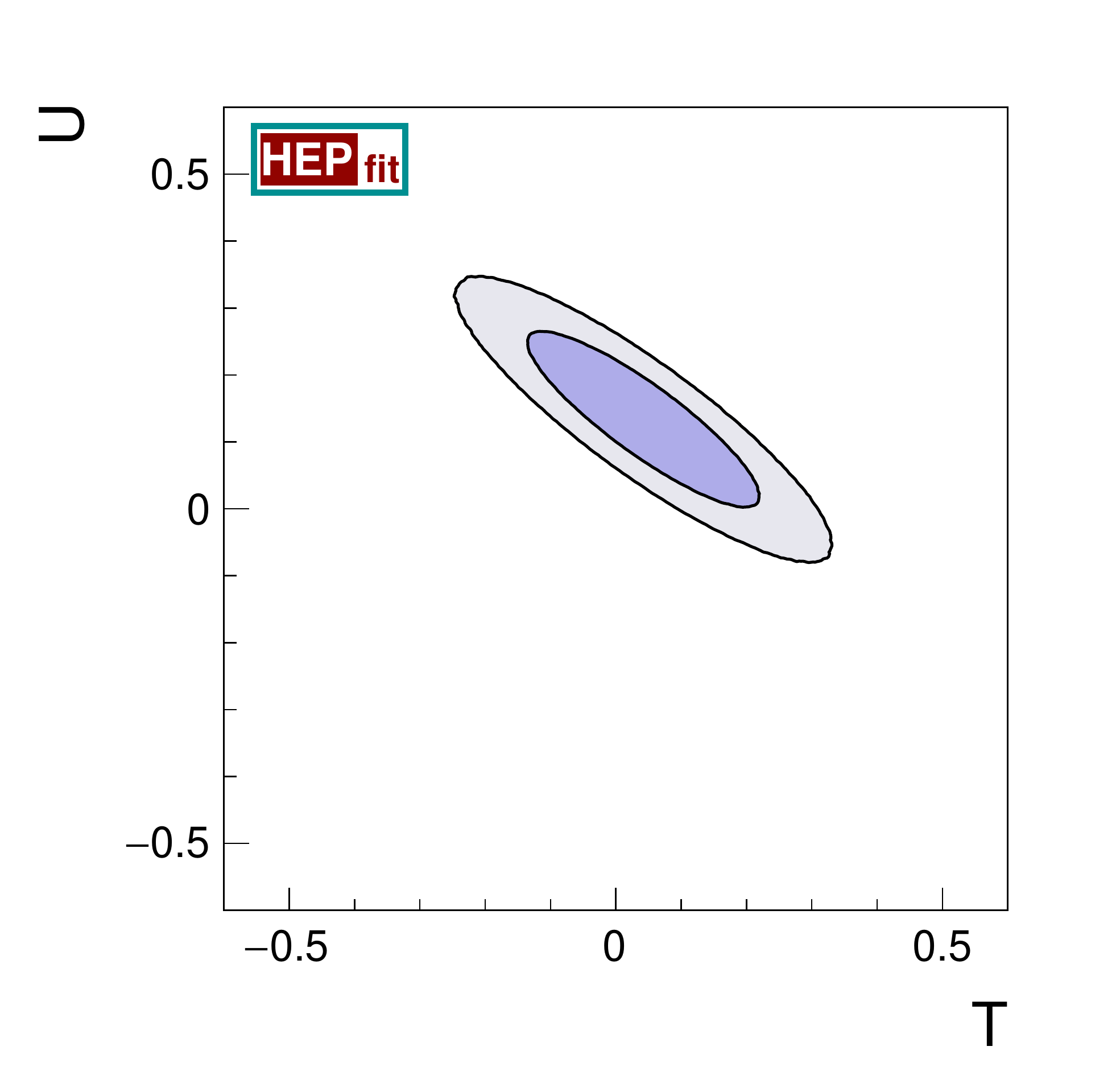}
    \caption{P.d.f's for oblique parameters from a global fit to all EWPO for the \emph{standard average} scenario. (Left panel) Scenario with $U=0$. (Center and right panels) Scenario with $U\not=0$.
    Dark (light) regions
    correspond to $68\%$ ($95\%$) probability ranges.}
    \label{fig:STU}
\end{figure*}

In view of the significant discrepancy between the SM prediction and the experimental average for $M_W$, we discuss next the implications of the new Tevatron result on scenarios of NP beyond the SM. In particular we discuss the case of NP models which mainly introduce sizable EW \textit{oblique} corrections (here denoted as \textit{oblique} models) and the case in which NP is described at the EW scale by more general effective interactions, taking as prototype example the dimension-six SM Effective Field Theory (SMEFT). 
Let us first consider a class of NP models in which the dominant contributions to EWPO are expected to arise as oblique corrections, i.e. via modifications of the EW gauge-boson self energies, and can thus be parameterized in terms of  the
$S$, $T$, and $U$ parameters introduced in
Ref.~\cite{Peskin:1990zt,Peskin:1991sw} (or equivalently by the
$\varepsilon_{1,2,3}$ parameters introduced in
refs.~\cite{Altarelli:1990zd,Altarelli:1991fk,Altarelli:1993sz}, although, for the sake of brevity, we consider here only the former set of parameters).  The explicit dependence of the EWPO on $S$, $T$,
and $U$ can be found in appendix A of Ref.~\cite{Ciuchini:2013pca}. If one assumes NP contributions to $U$ to be negligible, then a prediction for $M_W$ can be obtained from all other EWPO, as reported in Table \ref{tab:MW_fits}, and could reduce the SM discrepancy with the experimental value of $M_W$ to a tension at the $1.5~\!\sigma$ level. This scenario, $U \ll S,T$, is expected in extensions with heavy new physics where the SM gauge symmetries are realized linearly in the light fields, in which case $U$ is generated by interactions of mass dimension eight, and is then suppressed with respect to $S$ and $T$, which are given by dimension-six interactions.
Alternatively, to describe scenarios where sizable contributions to $U$ are generated, we also consider the case where this parameter is left free.~\footnote{The $STU$ results can also be used to derive constraints in terms of the three combinations of four dimension-six {\it oblique} operators that affect EWPO, namely $S,T,W$, and $Y$~\cite{Barbieri:2004qk}, via their relation with the $\varepsilon_{1,2,3}$ parameters~\cite{Barbieri:2004qk}.} 
In this case, since $U$ is only very loosely constrained by $\Gamma_W$, $M_W$ cannot be predicted with a reasonable accuracy. At the same time, this means that the apparent discrepancy with the new $M_W$ measurement can be solved by a nonvanishing $U$ parameter. In Tables \ref{tab:STU} and \ref{tab:STU_C} we report the results of a global fit, including $M_W$, for the oblique parameters, while the corresponding probability density functions (p.d.f.) are presented in Figs.~\ref{fig:STU} and \ref{fig:STU_C}. We also report the value of the {\it Information Criterion} (IC)  \cite{IC} of the fits, compared to the SM one. The posterior for the EWPO is reported in Tables \ref{tab:NP_fits} and \ref{tab:NP_fits_c}.

\begin{table}[htb]
    \centering
    \resizebox{0.47\textwidth}{!}{
    \begin{tabular}{c|c|rr|c|rrr}
 \hline
 & Result & \multicolumn{2}{c|}{Correlation} & Result & \multicolumn{3}{c}{Correlation} \\\hline
 & \multicolumn{3}{c|}{\scriptsize{(IC$_{\rm ST}$/IC$_{\rm SM}=25.0/80.2$)}} & \multicolumn{4}{c}{\scriptsize{(IC$_{\rm STU}$/IC$_{\rm SM}=25.3/80.2$)}} \\
 \hline 
$S$ & $ 0.100 \pm 0.073 $ & $1.00$ & & $ 0.005 \pm 0.096 $ & \phantom{+}$1.00$ & & \\ 
$T$ & $ 0.202 \pm 0.056 $ & $0.93$ & $1.00$ & $ 0.040 \pm 0.120 $ & $0.91$ & $1.00$ \\ 
$U$ & $-$ & $-$ & $-$ & $ 0.134 \pm 0.087 $ & $-0.65$ & $-0.88$ & $1.00$ \\ \hline
    \end{tabular}}
    \caption{Results of the global fit of the oblique parameters to all EWPO in the \emph{standard average} scenario.}
    \label{tab:STU}
\end{table}

\begin{table*}[htb]
    \centering
    \begin{tabular}{c|c|c|c|c}
\hline 
 & Measurement & ST & STU & SMEFT \\
 \hline
$M_W$ [GeV] & $ 80.4133 \pm 0.0080 $ & $ 80.4100 \pm 0.0077 $  & $ 80.4133 \pm 0.0080 $ & $ 80.4133 \pm 0.0080 $ \\ 
$\Gamma_{W}$ [GeV] & $ 2.085 \pm 0.042 $ & $ 2.09214 \pm 0.00072 $  & $ 2.09251 \pm 0.00075 $ & $ 2.0778 \pm 0.0070 $ \\ 
$\sin^2\theta_{\rm eff}^{\rm lept}(Q_{\rm FB}^{\rm had})$ & $ 0.2324 \pm 0.0012 $ & $ 0.23142 \pm 0.00013 $  & $ 0.23147 \pm 0.00014 $ & --\\ 
$P_{\tau}^{\rm pol}=\mathcal{A}_\ell$ & $ 0.1465 \pm 0.0033 $ & $ 0.1478 \pm 0.0011 $  & $ 0.1474 \pm 0.0011 $ & $ 0.1488 \pm 0.0014 $ \\ 
$\Gamma_{Z}$ [GeV] & $ 2.4955 \pm 0.0023 $ & $ 2.49812 \pm 0.00099 $  & $ 2.4951 \pm 0.0022 $ & $ 2.4955 \pm 0.0023 $ \\ 
$\sigma_{h}^{0}$ [nb] & $ 41.480 \pm 0.033 $ & $ 41.4910 \pm 0.0077 $  & $ 41.4905 \pm 0.0077 $ & $ 41.481 \pm 0.032 $ \\ 
$R^{0}_{\ell}$ & $ 20.767 \pm 0.025 $ & $ 20.7506 \pm 0.0084 $  & $ 20.7510 \pm 0.0084 $ & $ 20.769 \pm 0.024 $ \\ 
$A_{\rm FB}^{0, \ell}$ & $ 0.0171 \pm 0.0010 $ & $ 0.01638 \pm 0.00023 $  & $ 0.01630 \pm 0.00024 $ & $ 0.01659 \pm 0.00032 $ \\ 
$\mathcal{A}_{\ell}$ (SLD) & $ 0.1513 \pm 0.0021 $ & $ 0.1478 \pm 0.0011 $  & $ 0.1474 \pm 0.0011 $ & $ 0.1488 \pm 0.0014 $ \\ 
$R^{0}_{b}$ & $ 0.21629 \pm 0.00066 $ & $ 0.21591 \pm 0.00010 $  & $ 0.21591 \pm 0.00010 $ & $ 0.21632 \pm 0.00065 $ \\ 
$R^{0}_{c}$ & $ 0.1721 \pm 0.0030 $ & $ 0.172198 \pm 0.000054 $  & $ 0.172200 \pm 0.000054 $ & $ 0.17159 \pm 0.00099 $ \\ 
$A_{\rm FB}^{0, b}$ & $ 0.0996 \pm 0.0016 $ & $ 0.10362 \pm 0.00075 $  & $ 0.10336 \pm 0.00077 $ & $ 0.1008 \pm 0.0014 $ \\ 
$A_{\rm FB}^{0, c}$ & $ 0.0707 \pm 0.0035 $ & $ 0.07407 \pm 0.00058 $  & $ 0.07387 \pm 0.00059 $ & $ 0.0734 \pm 0.0022 $ \\ 
$\mathcal{A}_b$ & $ 0.923 \pm 0.020 $ & $ 0.934812 \pm 0.000097 $  & $ 0.934779 \pm 0.000099 $ & $ 0.903 \pm 0.013 $ \\ 
$\mathcal{A}_c$ & $ 0.670 \pm 0.027 $ & $ 0.66815 \pm 0.00052 $  & $ 0.66796 \pm 0.00053 $ & $ 0.658 \pm 0.020 $ \\ 
$\mathcal{A}_s$ & $ 0.895 \pm 0.091 $ & $ 0.935710 \pm 0.000096 $  & $ 0.935676 \pm 0.000097 $ & $ 0.905 \pm 0.012 $ \\ 
BR$_{W\to\ell\bar\nu_\ell}$ & $ 0.10860 \pm 0.00090 $ & $ 0.108386 \pm 0.000022 $  & $ 0.108380 \pm 0.000022 $ & $ 0.10900 \pm 0.00038 $ \\ 
$\sin^2\theta_{\rm eff}^{\rm lept}$ (HC) & $ 0.23143 \pm 0.00025 $ & $ 0.23142 \pm 0.00013 $  & $ 0.23147 \pm 0.00014 $ & --\\ 
$R_{uc}$ & $ 0.1660 \pm 0.0090 $ & $ 0.172220 \pm 0.000032 $  & $ 0.172222 \pm 0.000032 $ & $ 0.17161 \pm 0.00098 $ \\ \hline
    \end{tabular}
    \caption{Posterior distributions for the global fit to all EWPO in the NP scenarios discussed in the text. For the reader's convenience we also report experimental data in the first column. The measurements interpreted as determinations of the effective leptonic weak mixing angle, namely $\sin^2\theta_{\rm eff}^{\rm lept}(Q_{\rm FB}^{\rm had})$ and $\sin^2\theta_{\rm eff}^{\rm lept}$ (HC), are not included in the SMEFT fits.}
    \label{tab:NP_fits}
\end{table*}

We then relax the assumption of dominant oblique NP contributions and consider generic heavy NP within the formalism of the dimension-six SMEFT. Here we work in the so-called {\it Warsaw basis}~\cite{Grzadkowski:2010es} assuming fermion universality and, as in the fits presented above, we use the $\left\{\alpha,G_\mu,M_Z\right\}$ EW input scheme~\cite{Brivio:2021yjb}. In the Warsaw basis, there are a total of ten operators that can modify the EWPO at leading order, but only eight combinations of the corresponding Wilson coefficients can be constrained by the data in Table~\ref{tab:SM_std}~\cite{Falkowski:2014tna,Brivio:2017bnu}. 
Using the notation of~\cite{Grzadkowski:2010es}, these combinations can be written as, e.g.~\cite{Falkowski:2014tna} 
\begin{center}
\begin{align}
\hat{C}_{\varphi f}^{(1)}=&C_{\varphi f}^{(1)} - \frac{Y_f}{2} C_{\varphi D},~~~f=l,q,e,u,d,\\
\hat{C}_{\varphi f}^{(3)}=& C_{\varphi f}^{(3)} +\frac{c_w^2}{4s_w^2} C_{\varphi D} +\frac{c_w}{s_w} C_{\varphi WB},~~~f=l,q,\\
\hat{C}_{ll}=&\frac 12 ((C_{ll})_{1221}+(C_{ll})_{2112}) = (C_{ll})_{1221},
\label{eq:chat}
\end{align}
\end{center}
where $s_w$, $c_w$ are the sine and cosine of the weak mixing angle, $Y_f$ denotes the fermion hypercharge and we have absorbed the dependence on the cut-off scale of the SMEFT, $\Lambda$, in the Wilson coefficients, i.e. the above coefficients carry dimension of $[\mathrm{mass}]^{-2}$. 
Furthermore, the effective EW fermion couplings always depend on $\hat{C}_{ll}$ via the following combinations, fixed by the corresponding fermionic quantum numbers (see e.g. \cite{Azatov:2022kbs}), 
\begin{equation}
\hat{C}_{\varphi f}^{(3)}-\frac{c_w^2}{2s_w^2}\hat{C}_{ll}~~~~{\rm and}~~~~\hat{C}_{\varphi f}^{(1)}+Y_f\hat{C}_{ll},
\end{equation}
such that the effects of $\hat{C}_{ll}$ cannot be separated from other operators using only $Z$-pole observables. The flat direction can be broken by the $W$-boson mass, which depends on $\hat{C}_{\varphi l}^{(3)}-\hat{C}_{ll}/2$, or any observable sensitive to its value, e.g. the $W$-boson width $\Gamma_W$. The comparatively low precision of the experimental measurement of $\Gamma_W$ ($\sim 2\%$) thus results in a weak prediction for $M_W$ from the SMEFT fit, with an uncertainty somewhat below 2 GeV\footnote{This only accounts for the SMEFT parametric and SM intrinsic uncertainties but neglects the uncertainty associated to higher-order effects in the SMEFT, e.g. from dimension-eight contributions, which could be evaluated via the methods of \cite{Corbett:2021eux}.}, see Table~\ref{tab:MW_fits}, which can easily fit the experimental measurement, via a non-zero value of the combination $\hat{C}_{\varphi l}^{(3)}-\hat{C}_{ll}/2$. Indeed, as can be seen in Tables \ref{tab:SMEFT fit} and \ref{tab:SMEFT fit_c} for the standard and conservative scenarios, respectively, the two operators involved in the combination are strongly correlated between them, but also with $\hat{C}_{\varphi l}^{(1)}$. The latter correlation can be understood from the fact that the combination $\hat{C}_{\varphi l}^{(1)}+\hat{C}_{\varphi l}^{(3)}$ is the one that directly corrects the left-handed electron couplings, which is measured to the permil level. The extraction of this coupling from data, however, is typically correlated with the one on the right-handed coupling, sensitive to $\hat{C}_{\varphi e}$, complicating slightly more the correlation pattern in the output of the global fit. It is, in fact, in the information of the leptonic operators where one observes the main difference between the fits using the standard and conservative averages of the experimental values. This is reflected in changes in their correlations as well as mild changes, of order ten percent, in their uncertainties, whereas the central values of the Wilson coefficients stay approximately the same. The posterior for the EWPO in this case is also reported in Tables \ref{tab:NP_fits} and \ref{tab:NP_fits_c}.

\begin{table*}[htbp]
    \centering
    \begin{tabular}{c|c|rrrrrrrrr}
 \hline
 & Result & \multicolumn{8}{c}{Correlation Matrix} \\ 
 \hline 
 & \multicolumn{8}{c}{\scriptsize{(IC$_{\rm SMEFT}$/IC$_{\rm SM}=31.8/80.2$)}} \\
 \hline
$\hat{C}_{\varphi l}^{(1)}$ & $ -0.007 \pm 0.011 $      & $1.00$ & $ $ & $ $ & $ $ & $ $ & $ $ & $ $ & $ $ & $ $ \\ 
$\hat{C}_{\varphi l}^{(3)}$ & $ -0.042 \pm 0.015 $      & $-0.68$& $1.00$ & $ $ & $ $ & $ $ & $ $ & $ $ & $ $ & $ $  \\ 
$\hat{C}_{\varphi e}$       & $ -0.017 \pm 0.009 $      & $0.48$ & $0.04$ & $1.00$ & $ $ & $ $ & $ $ & $ $ & $ $  \\
$\hat{C}_{\varphi q}^{(1)}$ & $ -0.018\pm 0.044 $      & $-0.02$ & $-0.06$ & $-0.13$ & $1.00$ & $ $ & $ $ & $ $ & $ $ & $ $  \\
$\hat{C}_{\varphi q}^{(3)}$ & $ -0.113 \pm 0.043 $      & $-0.03$ & $0.04$ & $-0.16$ & $ -0.37$ & $1.00$& $ $ & $ $ & $ $ & $ $  \\
$\hat{C}_{\varphi u}$       & $ \phantom{+}0.090 \pm 0.150 $      & $0.06$ & $-0.04$ & $0.04$ & $0.61$ & $-0.77$ & $1.00$& $ $ & $ $ & $ $  \\
$\hat{C}_{\varphi d}$       & $ -0.630 \pm 0.250 $      & $-0.13$ & $-0.05$ & $-0.30$ & $0.40$ & $0.58$ & $-0.04$ & $1.00$& $ $ & $ $  \\
$\hat{C}_{ll}$              & $ -0.022 \pm 0.028 $     & $-0.80$ & $0.95$ & $-0.10$ & $-0.06$ & $-0.01$ & $-0.04$ & $-0.05$ & $1.00$ \\ \hline
%
%
%
    \end{tabular}
    \caption{Results from the dimension-six SMEFT fit in the {\it standard average} scenario. The values of the Wilson coefficients $\hat{C}_i$ are given in units of TeV$^{-2}$.}
    \label{tab:SMEFT fit}
\end{table*}

In conclusion, recent measurements of $m_t$~\cite{CMS-PAS-TOP-20-008} and $M_W$~\cite{CDF:2022hxs} are introducing some tensions in global fits of EW precision observables. In this Letter we have studied their impact on electroweak precision fits both in the SM and in some prototype scenarios of NP beyond the SM. Future EW precision measurements at both the LHC and the HL-LHC will add to this picture and contribute to confirm or resolve potential tensions in the SM.

\section*{Acknowledgements}

This work was supported in part by the Italian Ministry of Research (MIUR) under grant PRIN 20172LNEEZ.
The work of J.B. has been supported by the FEDER/Junta de Andaluc\'ia project grant P18-FRJ-3735. The work of L.R.  has been supported by the U.S. Department of Energy under grant DE-SC0010102.

\section*{Appendix on the \emph{conservative average} scenario}

\begin{figure*}[htb]
  \centering
  \includegraphics[width=.425\textwidth]{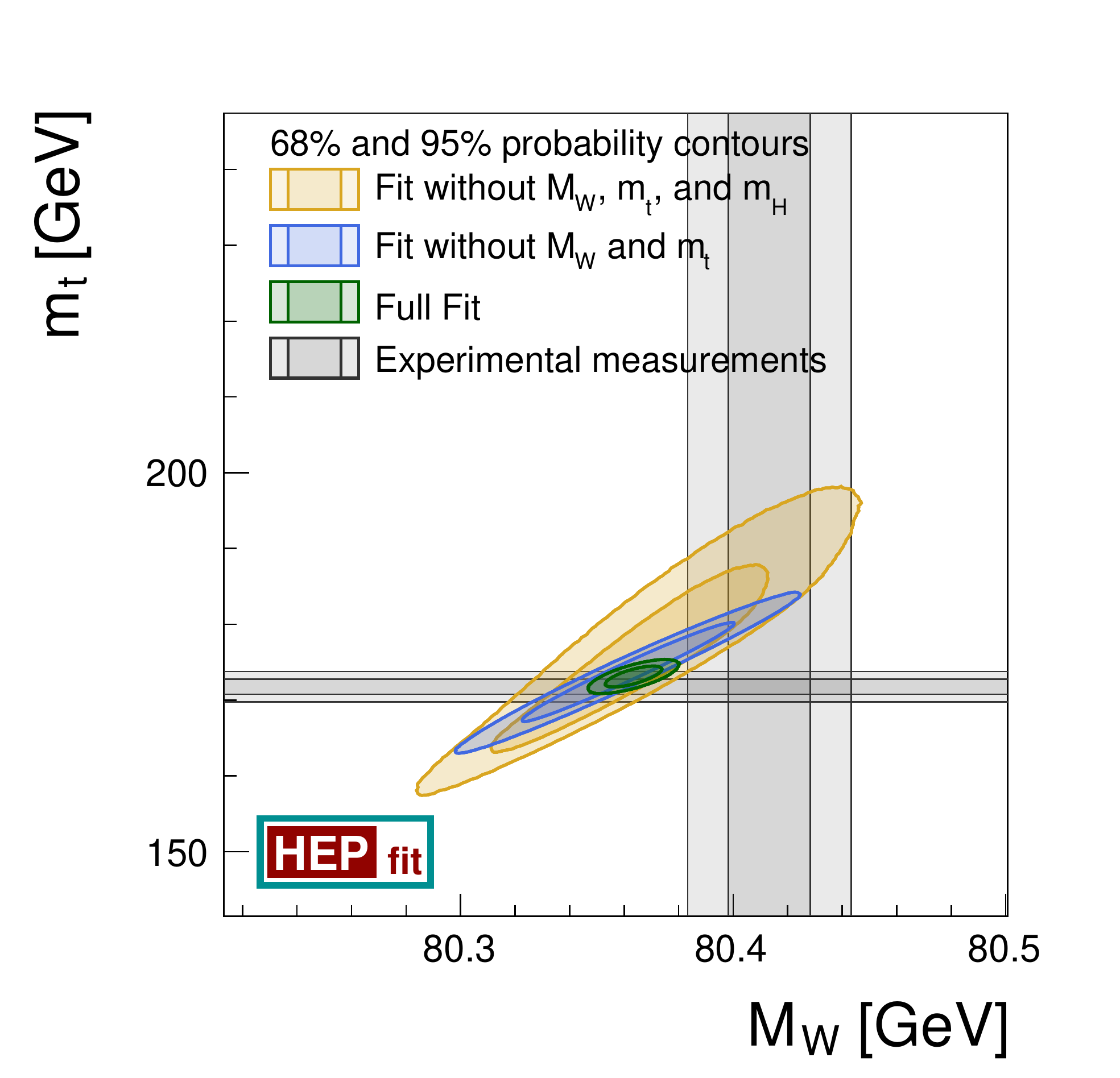}
  \includegraphics[width=.425\textwidth]{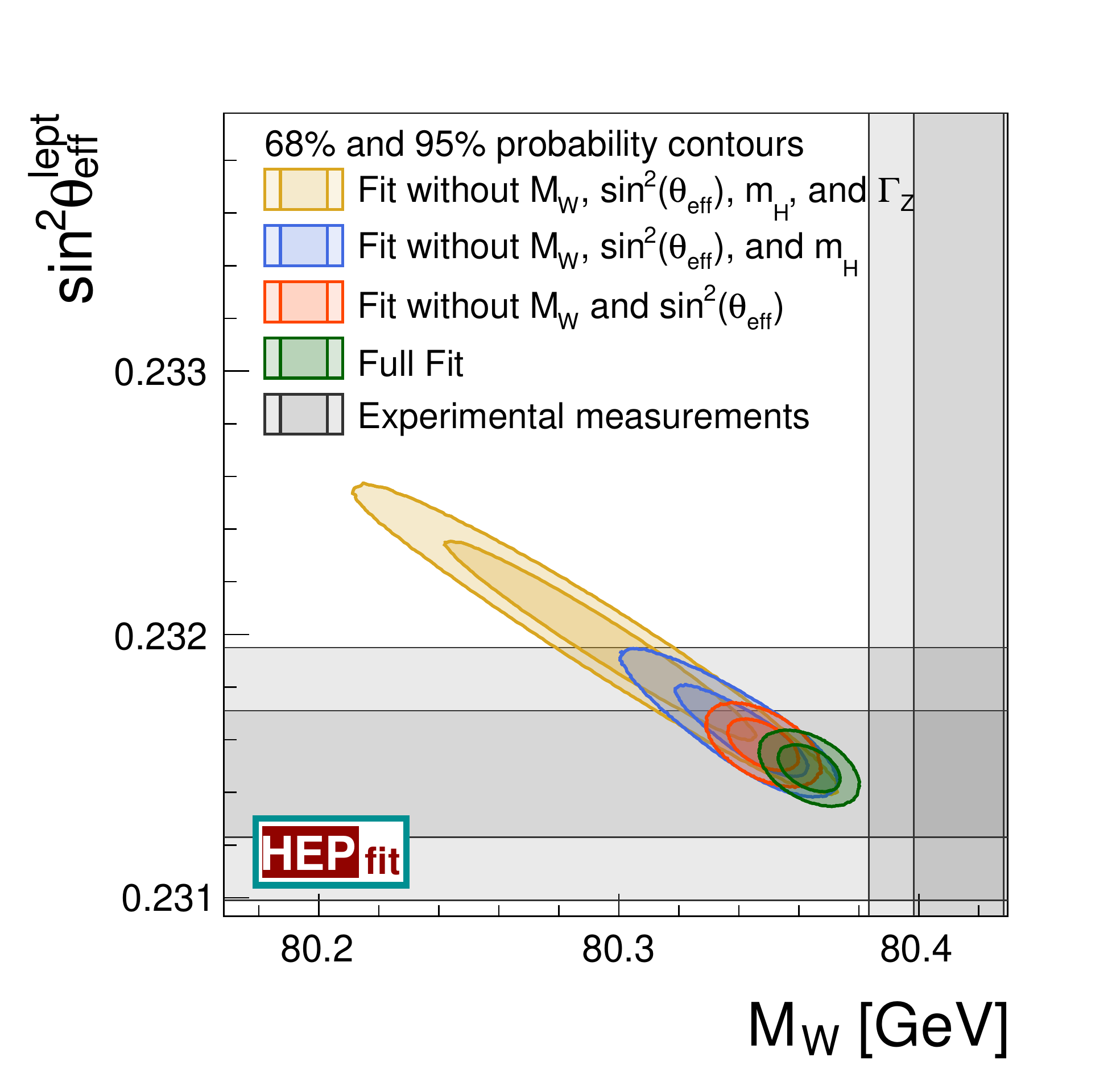}
  \caption{Same as Figure~\ref{fig:mtmwsm} in the \emph{conservative average} scenario.}
  \label{fig:mtmwsm_C}
\end{figure*}

\begin{figure*}[ht!]
    \centering
    \includegraphics[width=0.245\textwidth]{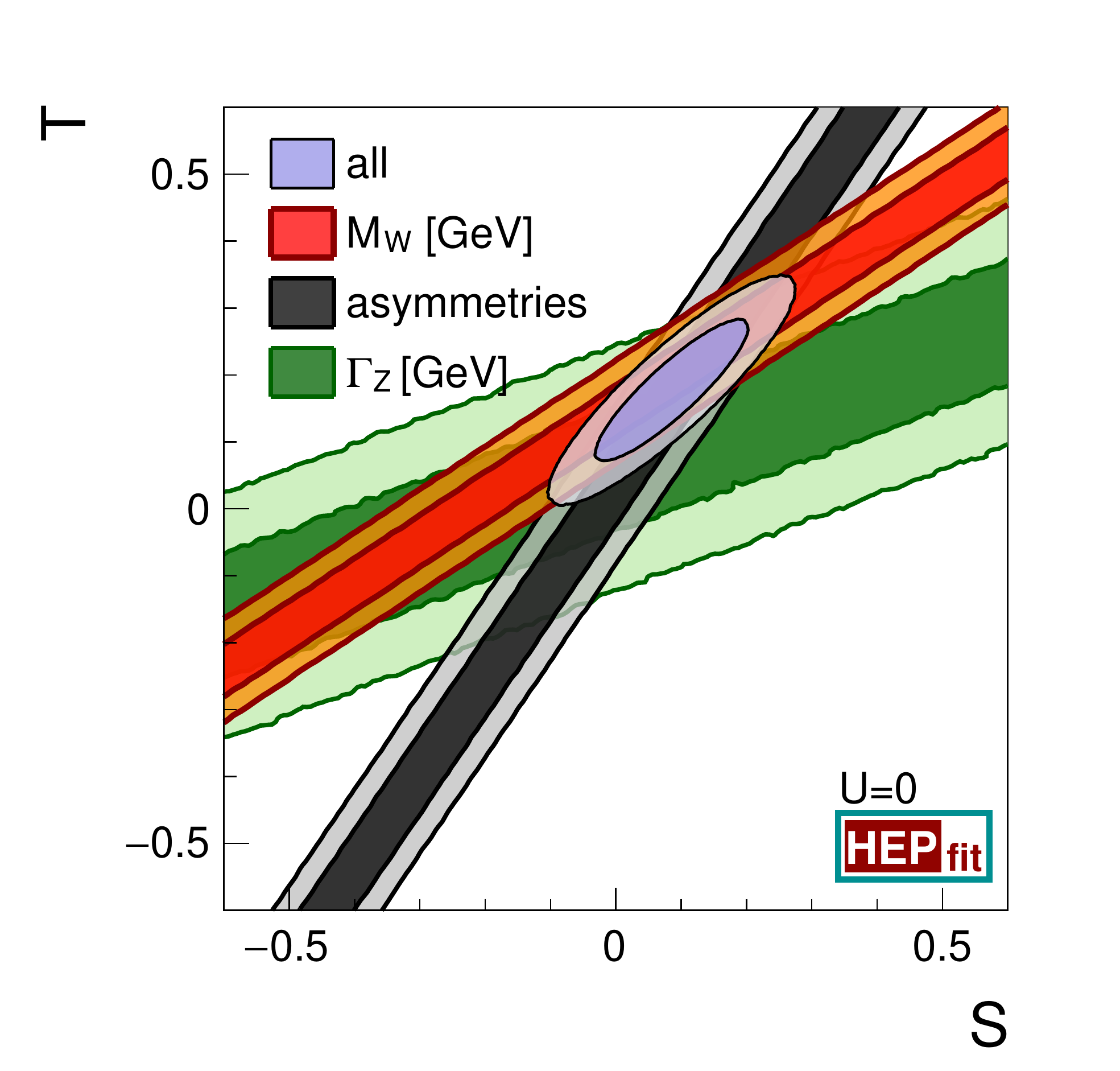}
    \includegraphics[width=0.245\textwidth]{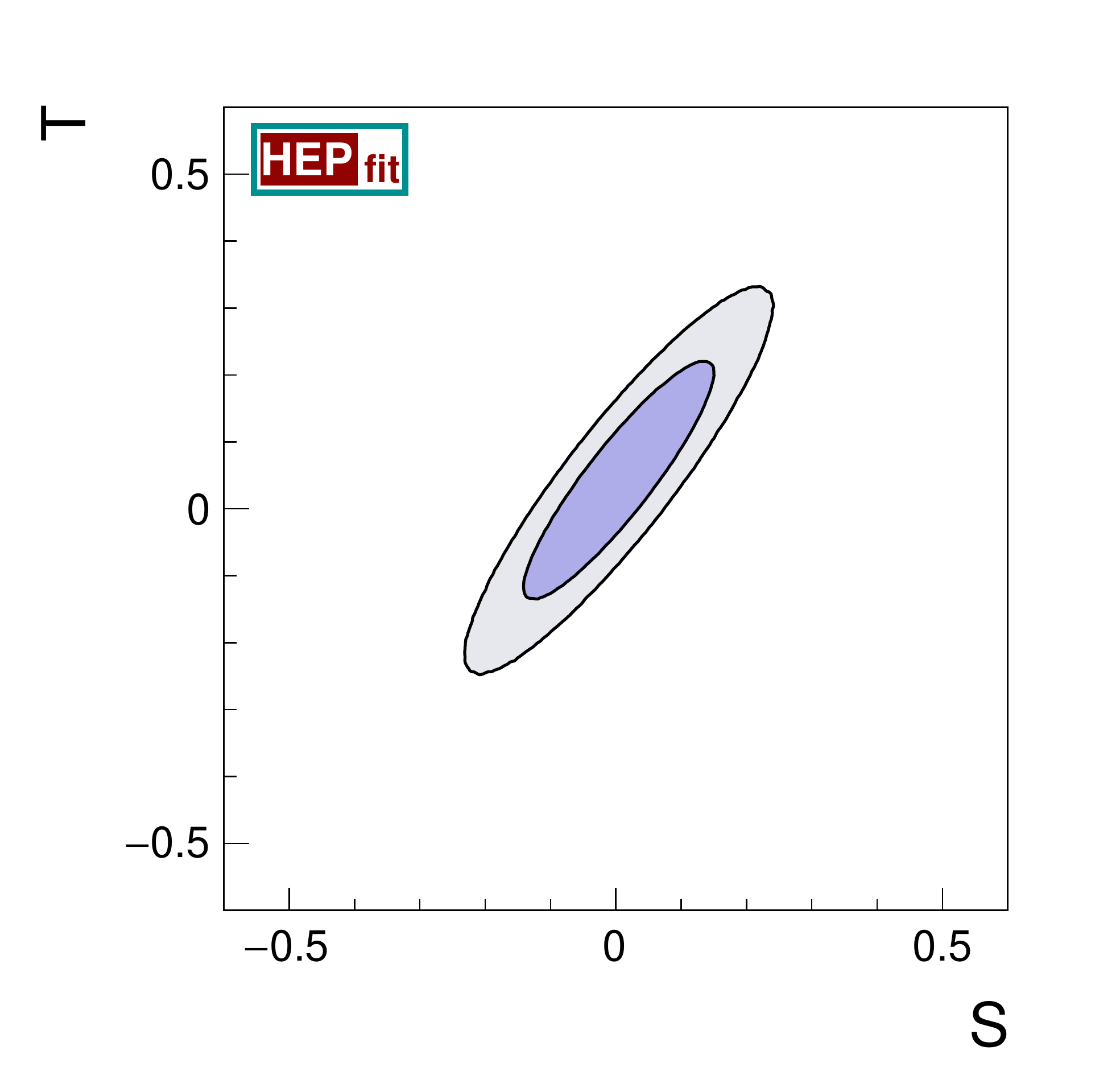}
    \includegraphics[width=0.245\textwidth]{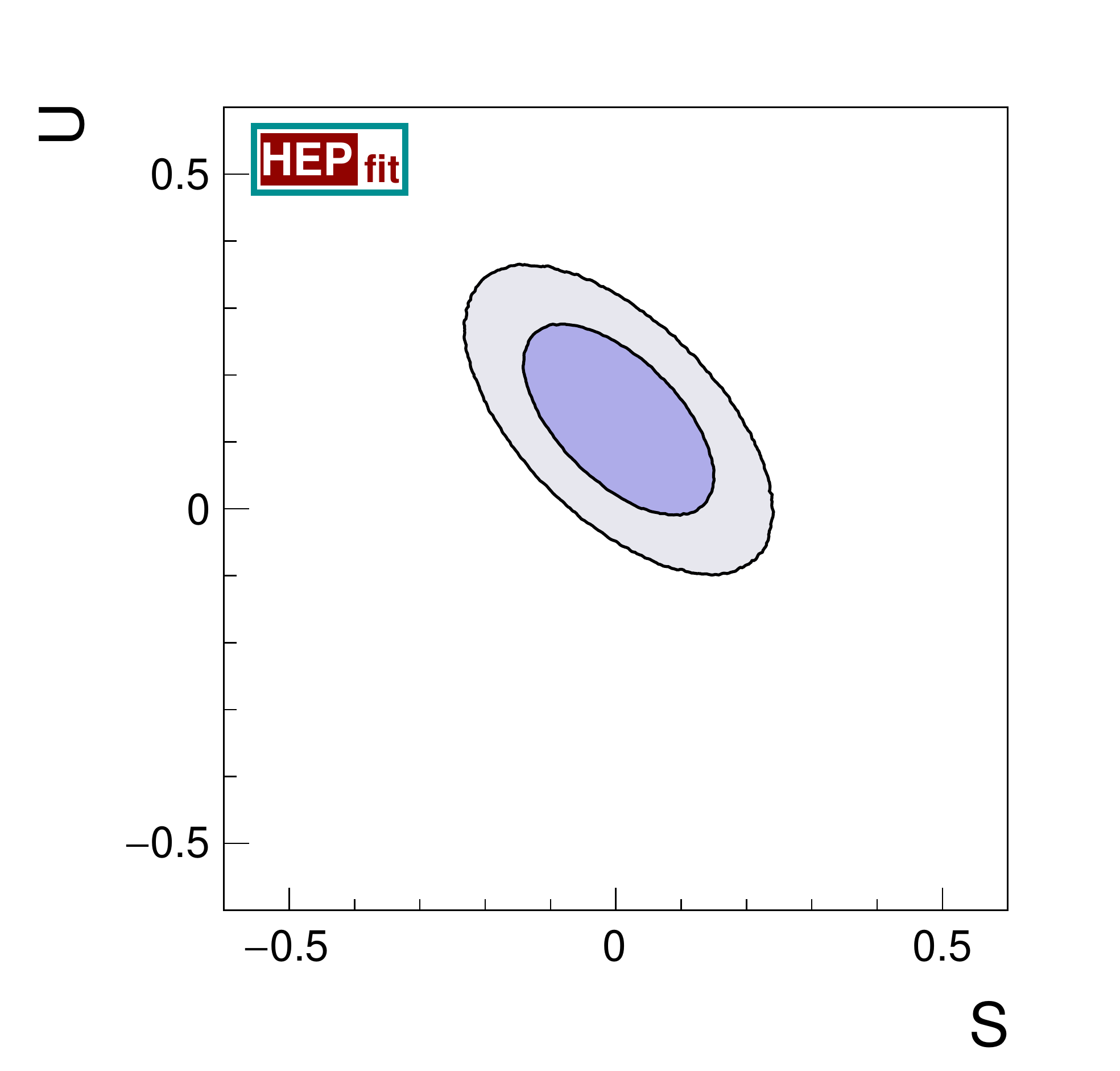}
    \includegraphics[width=0.245\textwidth]{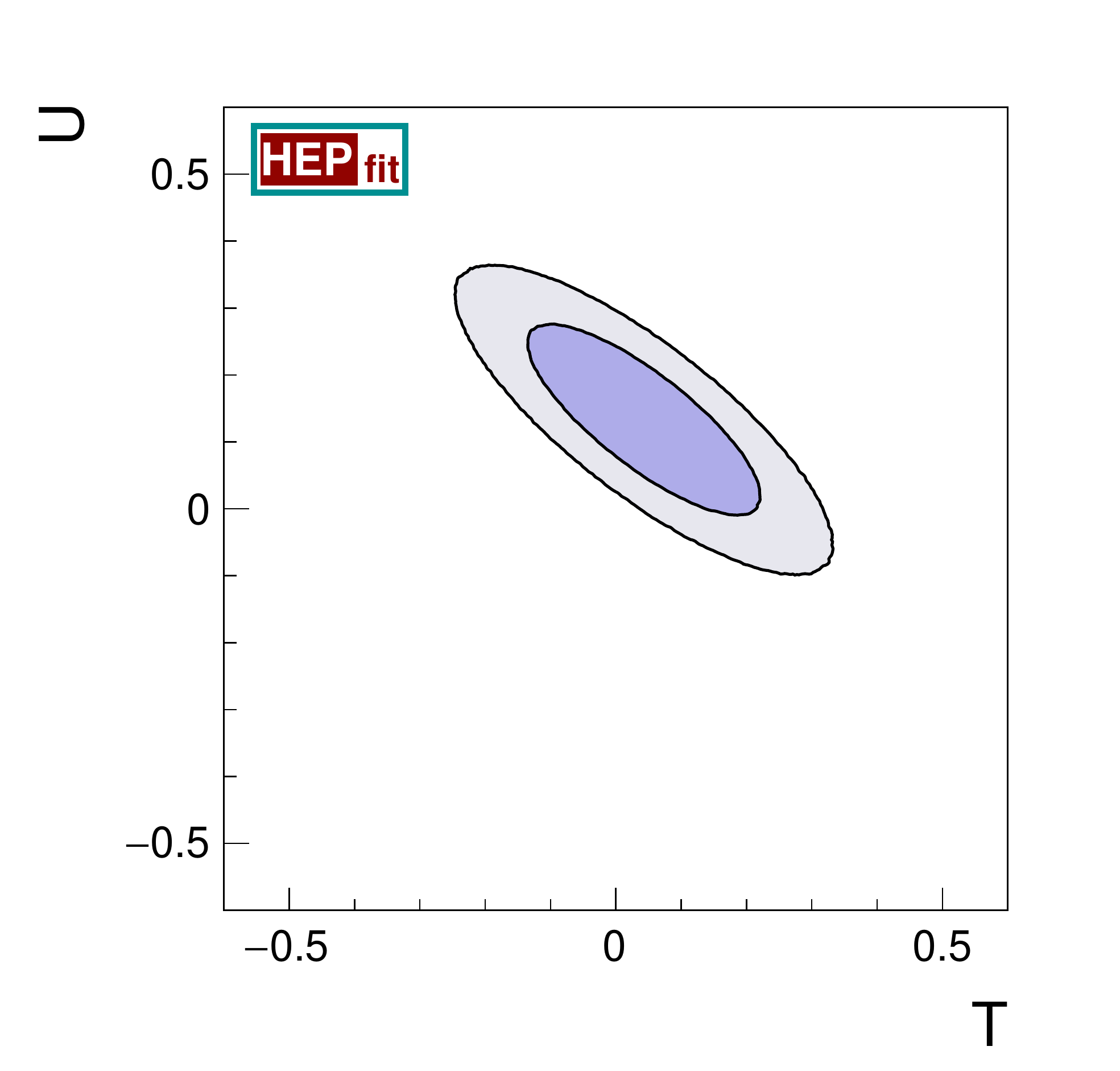}
    \caption{Same as Figure~\protect\ref{fig:STU} for the \emph{conservative average} scenario.}
    \label{fig:STU_C}
\end{figure*}

In this appendix we present the results of our analysis in the \emph{conservative average} scenario for $m_t$ and $M_W$. Figure \ref{fig:mtmwsm_C} presents the posteriors for different fits in the $m_t$ vs $M_W$ and $\sin^2\theta_{\rm eff}^{\rm lept}$ vs $M_W$ planes in the SM. Results of SM fits are reported in Table \ref{tab:SM_c}, while Figure~\ref{fig:STU_C} and Table~\ref{tab:STU_C} present results obtained in the scenario with dominant oblique NP contributions, and Table~\ref{tab:SMEFT fit_c} presents the corresponding results for the SMEFT. Posteriors for all EWPO in the NP scenarios considered are reported in Table~\ref{tab:NP_fits_c}.

\newpage

\begin{table*}[p!]
    \centering\resizebox{\textwidth}{!}{
    \begin{tabular}{l|c|c|cr|cr|cr}
\toprule
& Measurement & Posterior & Indirect/Prediction & Pull & Full Indirect & Pull & Full Prediction & Pull \\
\hline 
$\alpha_{s}(M_{Z})$ & $ 0.1177 \pm 0.0010 $ & $ 0.11786 \pm 0.00095 $ & $ 0.11930 \pm 0.00281 $ & $ -0.5 $ & $ 0.12174 \pm 0.00473 $ & $ -0.8 $ & $ 0.1177 \pm 0.0010 $ &$ - $ \\ 
& & $[ 0.11603 , 0.11972 ]$ & $[ 0.11371 , 0.12482 ]$ & & $[ 0.1126 , 0.1311 ]$ & & $[ 0.1157 , 0.1197 ]$  \\ 
$\Delta\alpha^{(5)}_{\mathrm{had}}(M_Z) $& $ 0.02766 \pm 0.00010 $ & $ 0.027614 \pm 0.000097 $ & $ 0.026895 \pm 0.000394 $ & $ 1.9 $ & $ 0.027987 \pm 0.000699 $ & $ -0.5 $ & $ 0.02766 \pm 0.00010 $ &$ - $ \\ 
& & $[ 0.027422 , 0.027804 ]$ & $[ 0.026123 , 0.027677 ]$ & & $[ 0.02661 , 0.02935 ]$ & & $[ 0.02747 , 0.02786 ]$  \\ 
$M_Z$ [GeV] & $ 91.1875 \pm 0.0021 $ & $ 91.1887 \pm 0.0021 $ & $ 91.2227 \pm 0.0105 $ & $ -3.3 $ & $ 91.2111 \pm 0.0390 $ & $ -0.6 $ & $ 91.1875 \pm 0.0021 $ &$ - $ \\ 
& & $[ 91.1847 , 91.1927 ]$ & $[ 91.2024 , 91.2434 ]$ & & $[ 91.135 , 91.289 ]$ & & $[ 91.1834 , 91.1916 ]$  \\ 
$m_t$ [GeV] & $ 171.8 \pm 1.0 $ & $ 173.12 \pm 0.92 $ & $ 180.10 \pm 2.25 $ & $ -3.3 $ & $ 187.16 \pm 9.83 $ & $ -1.6 $ & $ 171.8 \pm 1.0 $ &$ - $ \\ 
& & $[ 171.30 , 174.92 ]$ & $[ 175.66 , 184.55 ]$ & & $[ 167.9 , 206.4 ]$ & & $[ 169.8 , 173.8 ]$  \\ 
$m_H$ [GeV] & $ 125.21 \pm 0.12 $ & $ 125.21 \pm 0.12 $ & $ 102.19 \pm 9.79 $ & $ 1.9 $ & $ 245.25 \pm 125.35 $ & $ -0.9 $ & $ 125.21 \pm 0.12 $ &$ - $ \\ 
& & $[ 124.97 , 125.45 ]$ & $[ 87.01 , 127.30 ]$ & & $[ 98.1 , 640.4 ]$ & & $[ 124.97 , 125.45 ]$  \\ 
\hline 
$M_W$ [GeV] & $ 80.413 \pm 0.015 $ & $ 80.3634 \pm 0.0068 $ & $ 80.3505 \pm 0.0077 $ & $ 3.7 $ & $ 80.4116 \pm 0.0146 $ & $ 0.0 $ & $ 80.3497 \pm 0.0079 $ &$ 3.7 $ \\ 
& & $[ 80.3500 , 80.3769 ]$ & $[ 80.3355 , 80.3655 ]$ & & $[ 80.383 , 80.440 ]$ & & $[ 80.3342 , 80.3653 ]$  \\ 
\hline 
$\Gamma_{W}$ [GeV] & $ 2.085 \pm 0.042 $ & $ 2.08859 \pm 0.00066 $ & $ 2.08859 \pm 0.00066 $ & $ -0.1 $ & $ 2.09426 \pm 0.00245 $ & $ -0.2 $ & $ 2.08743 \pm 0.00073 $ &$ 0.0 $ \\ 
& & $[ 2.08731 , 2.08988 ]$ & $[ 2.08732 , 2.08988 ]$ & & $[ 2.0894 , 2.0990 ]$ & & $[ 2.08601 , 2.08889 ]$  \\ 
\hline 
$\sin^2\theta_{\rm eff}^{\rm lept}(Q_{\rm FB}^{\rm had})$ & $ 0.2324 \pm 0.0012 $ & $ 0.231491 \pm 0.000059 $ & $ 0.231490 \pm 0.000059 $ & $ 0.8 $ & $ 0.231461 \pm 0.000136 $ & $ 0.8 $ & $ 0.231558 \pm 0.000068 $ &$ 0.7 $ \\ 
& & $[ 0.231376 , 0.231608 ]$ & $[ 0.231374 , 0.231607 ]$ & & $[ 0.23119 , 0.23173 ]$ & & $[ 0.231426 , 0.231691 ]$  \\ 
\hline 
$P_{\tau}^{\rm pol}=\mathcal{A}_\ell$ & $ 0.1465 \pm 0.0033 $ & $ 0.14725 \pm 0.00046 $ & $ 0.14727 \pm 0.00047 $ & $ -0.2 $ & $ 0.14750 \pm 0.00108 $ & $ -0.3 $ & $ 0.14674 \pm 0.00053 $ &$ -0.1 $ \\ 
& & $[ 0.14634 , 0.14817 ]$ & $[ 0.14635 , 0.14820 ]$ & & $[ 0.1454 , 0.1496 ]$ & & $[ 0.14570 , 0.14779 ]$  \\ 
\hline 
$\Gamma_{Z}$ [GeV] & $ 2.4955 \pm 0.0023 $ & $ 2.49453 \pm 0.00066 $ & $ 2.49434 \pm 0.00070 $ & $ 0.5 $ & $ 2.49528 \pm 0.00205 $ & $ 0.1 $ & $ 2.49396 \pm 0.00072 $ &$ 0.6 $ \\ 
& & $[ 2.49324 , 2.49584 ]$ & $[ 2.49295 , 2.49572 ]$ & & $[ 2.4912 , 2.4993 ]$ & & $[ 2.49257 , 2.49538 ]$  \\ 
$\sigma_{h}^{0}$ [nb] & $ 41.480 \pm 0.033 $ & $ 41.4908 \pm 0.0077 $ & $ 41.4929 \pm 0.0080 $ & $ -0.4 $ & $ 41.4616 \pm 0.0304 $ & $ 0.4 $ & $ 41.4924 \pm 0.0080 $ &$ -0.4 $ \\ 
& & $[ 41.4757 , 41.5059 ]$ & $[ 41.4772 , 41.5087 ]$ & & $[ 41.402 , 41.522 ]$ & & $[ 41.4767 , 41.5083 ]$  \\ 
$R^{0}_{\ell}$ & $ 20.767 \pm 0.025 $ & $ 20.7491 \pm 0.0080 $ & $ 20.7458 \pm 0.0086 $ & $ 0.8 $ & $ 20.7589 \pm 0.0218 $ & $ 0.2 $ & $ 20.7470 \pm 0.0087 $ &$ 0.8 $ \\ 
& & $[ 20.7333 , 20.7649 ]$ & $[ 20.7287 , 20.7627 ]$ & & $[ 20.716 , 20.802 ]$ & & $[ 20.7297 , 20.7638 ]$  \\ 
$A_{\rm FB}^{0, \ell}$ & $ 0.0171 \pm 0.0010 $ & $ 0.01626 \pm 0.00010 $ & $ 0.01625 \pm 0.00010 $ & $ 0.8 $ & $ 0.01631 \pm 0.00024 $ & $ 0.8 $ & $ 0.01615 \pm 0.00012 $ &$ 1.0 $ \\ 
& & $[ 0.01606 , 0.01647 ]$ & $[ 0.01605 , 0.01646 ]$ & & $[ 0.01585 , 0.01679 ]$ & & $[ 0.01592 , 0.01638 ]$  \\ 
\hline 
$\mathcal{A}_{\ell}$ (SLD) & $ 0.1513 \pm 0.0021 $ & $ 0.14725 \pm 0.00046 $ & $ 0.14728 \pm 0.00049 $ & $ 1.9 $ & $ 0.14750 \pm 0.00108 $ & $ 1.6 $ & $ 0.14674 \pm 0.00053 $ &$ 2.1 $ \\ 
& & $[ 0.14634 , 0.14817 ]$ & $[ 0.14632 , 0.14824 ]$ & & $[ 0.1454 , 0.1496 ]$ & & $[ 0.14570 , 0.14779 ]$  \\ 
$R^{0}_{b}$ & $ 0.21629 \pm 0.00066 $ & $ 0.21587 \pm 0.00010 $ & $ 0.21586 \pm 0.00011 $ & $ 0.7 $ & $ 0.21542 \pm 0.00037 $ & $ 1.2 $ & $ 0.21591 \pm 0.00011 $ &$ 0.6 $ \\ 
& & $[ 0.21566 , 0.21607 ]$ & $[ 0.21565 , 0.21607 ]$ & & $[ 0.21467 , 0.21613 ]$ & & $[ 0.21570 , 0.21611 ]$  \\ 
$R^{0}_{c}$ & $ 0.1721 \pm 0.0030 $ & $ 0.172210 \pm 0.000054 $ & $ 0.172210 \pm 0.000054 $ & $ 0.0 $ & $ 0.172400 \pm 0.000185 $ & $ -0.1 $ & $ 0.172190 \pm 0.000055 $ &$ -0.1 $ \\ 
& & $[ 0.172102 , 0.172316 ]$ & $[ 0.172103 , 0.172317 ]$ & & $[ 0.17205 , 0.17277 ]$ & & $[ 0.172082 , 0.172297 ]$  \\ 
$A_{\rm FB}^{0, b}$ & $ 0.0996 \pm 0.0016 $ & $ 0.10324 \pm 0.00033 $ & $ 0.10325 \pm 0.00035 $ & $ -2.2 $ & $ 0.10338 \pm 0.00076 $ & $ -2.1 $ & $ 0.10287 \pm 0.00037 $ &$ -2.0 $ \\ 
& & $[ 0.10259 , 0.10388 ]$ & $[ 0.10258 , 0.10393 ]$ & & $[ 0.10188 , 0.10489 ]$ & & $[ 0.10214 , 0.10361 ]$  \\ 
$A_{\rm FB}^{0, c}$ & $ 0.0707 \pm 0.0035 $ & $ 0.07377 \pm 0.00024 $ & $ 0.07377 \pm 0.00026 $ & $ -0.9 $ & $ 0.07391 \pm 0.00059 $ & $ -0.9 $ & $ 0.07348 \pm 0.00028 $ &$ -0.8 $ \\ 
& & $[ 0.07328 , 0.07425 ]$ & $[ 0.07327 , 0.07428 ]$ & & $[ 0.07275 , 0.07507 ]$ & & $[ 0.07293 , 0.07403 ]$  \\ 
$\mathcal{A}_b$ & $ 0.923 \pm 0.020 $ & $ 0.934746 \pm 0.000040 $ & $ 0.934746 \pm 0.000040 $ & $ -0.6 $ & $ 0.934594 \pm 0.000169 $ & $ -0.6 $ & $ 0.934721 \pm 0.000041 $ &$ -0.6 $ \\ 
& & $[ 0.934668 , 0.934825 ]$ & $[ 0.934668 , 0.934826 ]$ & & $[ 0.93426 , 0.93492 ]$ & & $[ 0.934640 , 0.934802 ]$  \\ 
$\mathcal{A}_c$ & $ 0.670 \pm 0.027 $ & $ 0.66789 \pm 0.00023 $ & $ 0.66789 \pm 0.00023 $ & $ 0.1 $ & $ 0.66816 \pm 0.00054 $ & $ 0.1 $ & $ 0.66766 \pm 0.00024 $ &$ 0.1 $ \\ 
& & $[ 0.66743 , 0.66834 ]$ & $[ 0.66743 , 0.66835 ]$ & & $[ 0.66712 , 0.66922 ]$ & & $[ 0.66718 , 0.66814 ]$  \\ 
\hline 
$\mathcal{A}_s$ & $ 0.895 \pm 0.091 $ & $ 0.935663 \pm 0.000043 $ & $ 0.935663 \pm 0.000043 $ & $ -0.4 $ & $ 0.935714 \pm 0.000099 $ & $ -0.5 $ & $ 0.935622 \pm 0.000045 $ &$ -0.5 $ \\ 
& & $[ 0.935580 , 0.935746 ]$ & $[ 0.935580 , 0.935746 ]$ & & $[ 0.935522 , 0.935909 ]$ & & $[ 0.935533 , 0.935709 ]$  \\ 
BR$_{W\to\ell\bar\nu_\ell}$ & $ 0.10860 \pm 0.00090 $ & $ 0.108382 \pm 0.000022 $ & $ 0.108382 \pm 0.000022 $ & $ 0.2 $ & $ 0.108293 \pm 0.000110 $ & $ 0.3 $ & $ 0.108386 \pm 0.000023 $ &$ 0.2 $ \\ 
& & $[ 0.108339 , 0.108425 ]$ & $[ 0.108339 , 0.108425 ]$ & & $[ 0.10808 , 0.10851 ]$ & & $[ 0.108340 , 0.108432 ]$  \\ 
$\sin^2\theta_{\rm eff}^{\rm lept}$ (HC) & $ 0.23143 \pm 0.00025 $ & $ 0.231491 \pm 0.000059 $ & $ 0.231496 \pm 0.000061 $ & $ -0.2 $ & $ 0.231461 \pm 0.000136 $ & $ -0.1 $ & $ 0.231558 \pm 0.000068 $ &$ -0.5 $ \\ 
& & $[ 0.231376 , 0.231608 ]$ & $[ 0.231376 , 0.231616 ]$ & & $[ 0.23119 , 0.23173 ]$ & & $[ 0.231426 , 0.231691 ]$  \\ 
\hline 
$R_{uc}$ & $ 0.1660 \pm 0.0090 $ & $ 0.172231 \pm 0.000033 $ & $ 0.172231 \pm 0.000033 $ & $ -0.7 $ & $ 0.172424 \pm 0.000180 $ & $ -0.7 $ & $ 0.172211 \pm 0.000034 $ &$ -0.7 $ \\ 
& & $[ 0.172167 , 0.172295 ]$ & $[ 0.172168 , 0.172296 ]$ & & $[ 0.17208 , 0.17279 ]$ & & $[ 0.172145 , 0.172277 ]$  \\ 
 \bottomrule
    \end{tabular}}
    \caption{Same as Table \protect\ref{tab:SM_std} in the \emph{conservative average} scenario.}
    \label{tab:SM_c}
%
\vspace{1.5cm}
%
    \centering
    \resizebox{0.47\textwidth}{!}{
   \begin{tabular}{c|c|rr|c|rrr}
 \hline
 & Result & \multicolumn{2}{c|}{Correlation} & Result & \multicolumn{3}{c}{Correlation} \\\hline
 & \multicolumn{3}{c|}{\scriptsize{(IC$_{\rm ST}$/IC$_{\rm SM}=24.5/37.1$)}} & \multicolumn{4}{c}{\scriptsize{(IC$_{\rm STU}$/IC$_{\rm SM}=25.3/37.1$)}} \\
 \hline 
$S$ & $ 0.086 \pm 0.077 $ & $1.00$ & & $ 0.004 \pm 0.096 $ & $1.00$ & & \\ 
$T$ & $ 0.177 \pm 0.070 $ & $0.89$ & $1.00$ & $ 0.040 \pm 0.120 $ & $0.90$ & $1.00$ & \\ 
$U$ & $-$ & $-$ & $-$ & $ 0.134 \pm 0.095 $ & $-0.60$ & $-0.81$ & $1.00$ \\ \hline
    \end{tabular}}
    \caption{Same as Table~\protect\ref{tab:STU} in the \emph{conservative average} scenario.}
    \label{tab:STU_C}
\end{table*}

\begin{table*}[p]
    \centering
    \begin{tabular}{c|c|rrrrrrrrr}
 \hline
 & Result & \multicolumn{8}{c}{Correlation Matrix} \\ 
 \hline 
 & \multicolumn{8}{c}{\scriptsize{(IC$_{\rm SMEFT}$/IC$_{\rm SM}=32.0/37.1$)}} \\ 
 \hline
$\hat{C}_{\varphi l}^{(1)}$ & $ -0.007\pm 0.012 $  & $1.00$ & $ $ & $ $ & $ $ & $ $ & $ $ & $ $ & $ $ & $ $ \\ 
$\hat{C}_{\varphi l}^{(3)}$ & $ -0.042 \pm 0.018 $ & $-0.44$& $1.00$ & $ $ & $ $ & $ $ & $ $ & $ $ & $ $ & $ $  \\ 
$\hat{C}_{\varphi e}$       & $ -0.017 \pm 0.010 $ & $0.52$ & $0.31$ & $1.00$ & $ $ & $ $ & $ $ & $ $ & $ $ & $ $  \\
$\hat{C}_{\varphi q}^{(1)}$ & $ -0.018 \pm 0.045 $ & $-0.02$ & $-0.05$ & $-0.12$ & $1.00$ & $ $ & $ $ & $ $ & $ $ & $ $  \\
$\hat{C}_{\varphi q}^{(3)}$ & $ -0.114 \pm 0.044 $ & $0.02$ & $0.14$ & $-0.02$ & $-0.36$ & $1.00$& $ $ & $ $ & $ $ & $ $  \\
$\hat{C}_{\varphi u}$       & $ \phantom{+}0.090 \pm 0.150 $ & $0.05$ & $-0.04$ & $0.02$ & $0.61$ & $-0.76$ & $1.00$& $ $ & $ $ & $ $  \\
$\hat{C}_{\varphi d}$       & $ -0.630 \pm 0.250 $ & $-0.13$ & $-0.04$ & $-0.25$ & $0.40$ & $0.57$ & $-0.04$ & $1.00$& $ $ & $ $  \\
$\hat{C}_{ll}$              & $ -0.022 \pm 0.028 $ & $-0.72$ & $0.89$ & $0.01$ & $-0.06$ & $0.03$ & $-0.04$ & $-0.05$ & $1.00$ \\ \hline
    \end{tabular}
    \caption{Same as Table~\protect\ref{tab:SMEFT fit} for the \emph{conservative average} scenario.}
    \label{tab:SMEFT fit_c}
%
\vspace{2.5cm}
%
    \centering
    \begin{tabular}{c|c|c|c|c}
\hline 
 & Measurement & ST & STU & SMEFT \\
 \hline
$M_W$ [GeV] & $ 80.413 \pm 0.015 $ & $ 80.403 \pm 0.013 $  & $ 80.413 \pm 0.015 $ & $ 80.413 \pm 0.015 $ \\ 
$\Gamma_{W}$ [GeV] & $ 2.085 \pm 0.042 $ & $ 2.0916 \pm 0.0011 $  & $ 2.0925 \pm 0.0012 $ & $ 2.0778 \pm 0.0070 $ \\ 
$\sin^2\theta_{\rm eff}^{\rm lept}(Q_{\rm FB}^{\rm had})$ & $ 0.2324 \pm 0.0012 $ & $ 0.23143 \pm 0.00014 $  & $ 0.23147 \pm 0.00014 $ & -- \\ 
$P_{\tau}^{\rm pol}=\mathcal{A}_\ell$ & $ 0.1465 \pm 0.0033 $ & $ 0.1478 \pm 0.0011 $  & $ 0.1474 \pm 0.0011 $ & $ 0.1488 \pm 0.0014 $ \\ 
$\Gamma_{Z}$ [GeV] & $ 2.4955 \pm 0.0023 $ & $ 2.4976 \pm 0.0012 $  & $ 2.4951 \pm 0.0022 $ & $ 2.4955 \pm 0.0023 $ \\ 
$\sigma_{h}^{0}$ [nb] & $ 41.480 \pm 0.033 $ & $ 41.4909 \pm 0.0077 $  & $ 41.4905 \pm 0.0077 $ & $ 41.482 \pm 0.033 $ \\ 
$R^{0}_{\ell}$ & $ 20.767 \pm 0.025 $ & $ 20.7507 \pm 0.0084 $  & $ 20.7512 \pm 0.0084 $ & $ 20.769 \pm 0.025 $ \\ 
$A_{\rm FB}^{0, \ell}$ & $ 0.0171 \pm 0.0010 $ & $ 0.01637 \pm 0.00023 $  & $ 0.01630 \pm 0.00024 $ & $ 0.01660 \pm 0.00032 $ \\ 
$\mathcal{A}_{\ell}$ (SLD) & $ 0.1513 \pm 0.0021 $ & $ 0.1478 \pm 0.0011 $  & $ 0.1474 \pm 0.0011 $ & $ 0.1488 \pm 0.0014 $ \\ 
$R^{0}_{b}$ & $ 0.21629 \pm 0.00066 $ & $ 0.21591 \pm 0.00011 $  & $ 0.21591 \pm 0.00011 $ & $ 0.21632 \pm 0.00065 $ \\ 
$R^{0}_{c}$ & $ 0.1721 \pm 0.0030 $ & $ 0.172199 \pm 0.000055 $  & $ 0.172199 \pm 0.000055 $ & $ 0.17160 \pm 0.00099 $ \\ 
$A_{\rm FB}^{0, b}$ & $ 0.0996 \pm 0.0016 $ & $ 0.10359 \pm 0.00075 $  & $ 0.10337 \pm 0.00077 $ & $ 0.1009 \pm 0.0014 $ \\ 
$A_{\rm FB}^{0, c}$ & $ 0.0707 \pm 0.0035 $ & $ 0.07403 \pm 0.00059 $  & $ 0.07385 \pm 0.00059 $ & $ 0.0735 \pm 0.0022 $ \\ 
$\mathcal{A}_b$ & $ 0.923 \pm 0.020 $ & $ 0.934807 \pm 0.000097 $  & $ 0.934779 \pm 0.000100 $ & $ 0.903 \pm 0.013 $ \\ 
$\mathcal{A}_c$ & $ 0.670 \pm 0.027 $ & $ 0.66811 \pm 0.00052 $  & $ 0.66797 \pm 0.00053 $ & $ 0.658 \pm 0.020 $ \\ 
$\mathcal{A}_s$ & $ 0.895 \pm 0.091 $ & $ 0.935705 \pm 0.000096 $  & $ 0.935677 \pm 0.000097 $ & $ 0.905 \pm 0.012 $ \\ 
BR$_{W\to\ell\bar\nu_\ell}$ & $ 0.10860 \pm 0.00090 $ & $ 0.108385 \pm 0.000022 $  & $ 0.108380 \pm 0.000022 $ & $ 0.10900 \pm 0.00038 $ \\ 
$\sin^2\theta_{\rm eff}^{\rm lept}$ (HC) & $ 0.23143 \pm 0.00025 $ & $ 0.23143 \pm 0.00014 $  & $ 0.23147 \pm 0.00014 $ & -- \\ 
$R_{uc}$ & $ 0.1660 \pm 0.0090 $ & $ 0.172221 \pm 0.000034 $  & $ 0.172221 \pm 0.000034 $ & $ 0.17162 \pm 0.00099 $ \\ \hline
    \end{tabular}
    \caption{Same as Table~\protect\ref{tab:NP_fits} for the \emph{conservative average} scenario.} 
    \label{tab:NP_fits_c}
\end{table*}

\clearpage

\bibliography{hepbiblio}
 
\end{document}